\documentclass[twocolumn,aps,prb,groupedaddress,showkeys,showpacs]{revtex4}
\usepackage{bm}
\usepackage{graphicx}
\catcode`@=11
\def\matrix#1{\null\,\vcenter{\normalbaselines\m@th
    \ialign{\hfil$##$\hfil&&\quad\hfil$##$\hfil\crcr
      \mathstrut\crcr\noalign{\kern-\baselineskip}
      #1\crcr\mathstrut\crcr\noalign{\kern-\baselineskip}}}\,}
\def\pmatrix#1{\left(\matrix{#1}\right)}
\def\eqalign#1{\null\,\vcenter{\openup\jot\m@th
\ialign{\strut\hfil$\displaystyle{##}$&$\displaystyle{{}##}$\hfil
\crcr#1\crcr}}\,}
\def\Journal #1,#2,#3,#4#5#6#7{#1 {\bf #2}, #3 (#4#5#6#7)}
\def\lsim{\lower -0.3ex \hbox{$<$} \kern -0.75em \lower 0.7ex \hbox{$\sim$}}
\def\gsim{\lower -0.3ex \hbox{$>$} \kern -0.75em \lower 0.7ex \hbox{$\sim$}}
\def\i{i}
\catcode`@=12
\begin{document}
%
\title{Transmission through a boundary between monolayer and bilayer graphene}
\author{Takeshi Nakanishi$^{1}$, Mikito Koshino$^2$, and Tsuneya Ando$^2$}
\affiliation{$^1$Nanotube Research Center, AIST, 1--1--1 Higashi, Tsukuba 305-8565, Japan\\
$^2$Department of Physics, Tokyo Institute of Technology, 2--12--1 Ookayama, Meguro-ku, Tokyo 152-8551, Japan}
%
\begin{abstract}
%
The electron transmission between monolayer and bilayer graphene is theoretically studied for zigzag and armchair boundaries within an effective-mass scheme.
Due to the presence of an evanescent wave in the bilayer graphene, traveling modes are well connected to each other.
The transmission through the boundary is strongly dependent on the incident angle and the dependence is opposite between the K and K' points, leading to valley polarization of transmitted wave.
\par
%
\end{abstract}
\maketitle
%
\section{Introduction} \label{Sec:Introduction}
%
Graphene, the latest addition to the family of two-dimensional materials, is distinguished by its unusual electron dynamics governed by the Dirac equation.\cite{McClure_1956a,Slonczewski_and_Weiss_1958a,Ando_2005a,Ando_2007d}
Wave functions are characterized by spinor whose orientation is inextricably linked to the direction of the electron momentum in a different manner between monolayer and bilayer graphenes.\cite{Ando_et_al_1998b,Novoselov_et_al_2006a,McCann_and_Falko_2006a}
Recently monolayer and bilayer graphenes were fabricated using the method of mechanical exfoliation\cite{Novoselov_et_al_2006a,Novoselov_et_al_2004a} and epitaxially.\cite{Berger_et_al_2004a,Ohta_et_al_2006a}
The purpose of this paper is to study the electron transmission through boundary between monolayer and bilayer graphenes and show that strong valley polarization is induced in the transmission probability through the boundary.
\par
%
Transport properties in a monolayer graphene are quite intriguing, and the conductivity with/without a magnetic field including the Hall effect,\cite{Shon_and_Ando_1998a,Zheng_and_Ando_2002a} quantum corrections to the conductivity,\cite{Suzuura_and_Ando_2002b} and the dynamical transport\cite{Ando_et_al_2002a} were theoretically investigated prior to experiments.
The magnetotransport was measured including the integer quantum Hall effect, demonstrating the validity of the neutrino description of the electronic states.\cite{Novoselov_et_al_2005a,Zhang_et_al_2005a}
Bilayer graphene composed of a pair of graphene layers\cite{Novoselov_et_al_2006a,Ohta_et_al_2006a,Castro_et_al_2007a,Oostinga_et_al_2008a} has a zero-gap structure with quadratic dispersion different from a linear dispersion in a monolayer graphene.\cite{McCann_and_Falko_2006a,Koshino_and_Ando_2006a,Katsnelson_2006b,McCann_2006a,Guinea_et_al_2006a,Snyman_and_Beenakker_2007a,Koshino_2009a,San-Jose_et_al_2009a}
\par
%
In graphenes, states associated with K and K' points or valleys are degenerate.
A possible lifting of the degeneracy has been experimentally observed in high magnetic fields,\cite{Zhang_et_al_2006a} and there have been various suggestions on mechanisms leading to valley splitting and/or polarization.\cite{Koshino_and_Ando_2007a,Recher_et_al_2007a,Pereira_and_Schulz_2008a,Pereira_et_al_2009a,Nomura_and_MacDonald_2006a,Shibata_and_Nomura_2008a,Koshino_and_McCann_2010a}
A way to detect valley polarization is proposed with the use of a superconducting contact.\cite{Akhmerov_and_Beenakker_2007a}
\par
%
In a graphene sheet with a finite width, localized edge states are formed, when the boundary is in a certain specific direction.\cite{Fujita_et_al_1996a,Nakada_et_al_1996a}
Edge states of monolayer graphene ribbons have been a subject of extensive theoretical study.\cite{Wakabayashi_et_al_1999a,Wakabayashi_and_Sigrist_2000a,Wakabayashi_2001a,Wakabayashi_2002a,McCann_and_Falko_2004a,Brey_and_Fertig_2006b,Peres_et_al_2006b,Son_et_al_2006b,Son_et_al_2006c,Obradovic_et_al_2006a,Yang_et_al_2007a,Yang_et_al_2008b,Li_and_Lu_2008a,Raza_and_Kan_2008a,Ryzhii_et_al_2008a,Wassmann_et_al_2008a,Nguyen_et_al_2009a,Gunlycke_and_White_2010a}
The electron transport along the boundary has been calculated and characterized by odd number of channels in each valley.\cite{Wakabayashi_2002a}
When the number of occupied subbands is odd, a perfectly conducting channel transmitting through the ribbon is present\cite{Takane_2004c,Takane_and_Wakabayashi_2007a,Wakabayashi_et_al_2007a,Wakabayashi_et_al_2009a,Kobayashi_et_al_2009a} as in metallic carbon nanotubes.\cite{Ando_et_al_1998b,Ando_and_Nakanishi_1998a,Ando_and_Suzuura_2002a}
A way to make valley filtering has been proposed with the explicit use of the fact that only a single right- and left-going wave can carry current at each of the K and K' points.\cite{Rycerz_et_al_2007a} 
Recently, edge states in bilayer graphene were studied\cite{Castro_et_al_2008a,Sahu_et_al_2008a} and conductance through quantum structures consisting of monolayer and bilayer graphenes were calculated.\cite{Nilsson_et_al_2007a,Gonzalez_et_al_2010a}
\par
%
In this paper we study boundary conditions between monolayer and bilayer graphenes and calculate the transmission probability as a function of the electron concentration and the incident angle of injected wave.
In Sec.\ II the treatment of electronic states in a {\bf k}$\cdot${\bf p} scheme is briefly reviewed and boundary conditions are derived in Sec.\ III.
Valley polarization is shown in Sec.\ IV under the condition that the electron density in both monolayer and bilayer regions is the same.
Numerical results are presented in Sec.\ V and discussion and short summary are given in Sec.\ VI.
Analytic results in the vicinity of the Dirac point for zigzag and armchair boundaries are discussed in Appendix \ref{Sec:Low_Energy_Approximation} and \ref{Sec:Armchair_Boundary}, respectively, and the number of edges states localized at boundaries is discussed in Appendix \ref{Sec:Edge_States}.
\par
%
\begin{figure*}[!t]
$$
\begin{array}{ccc}
\includegraphics[width=6.0cm]{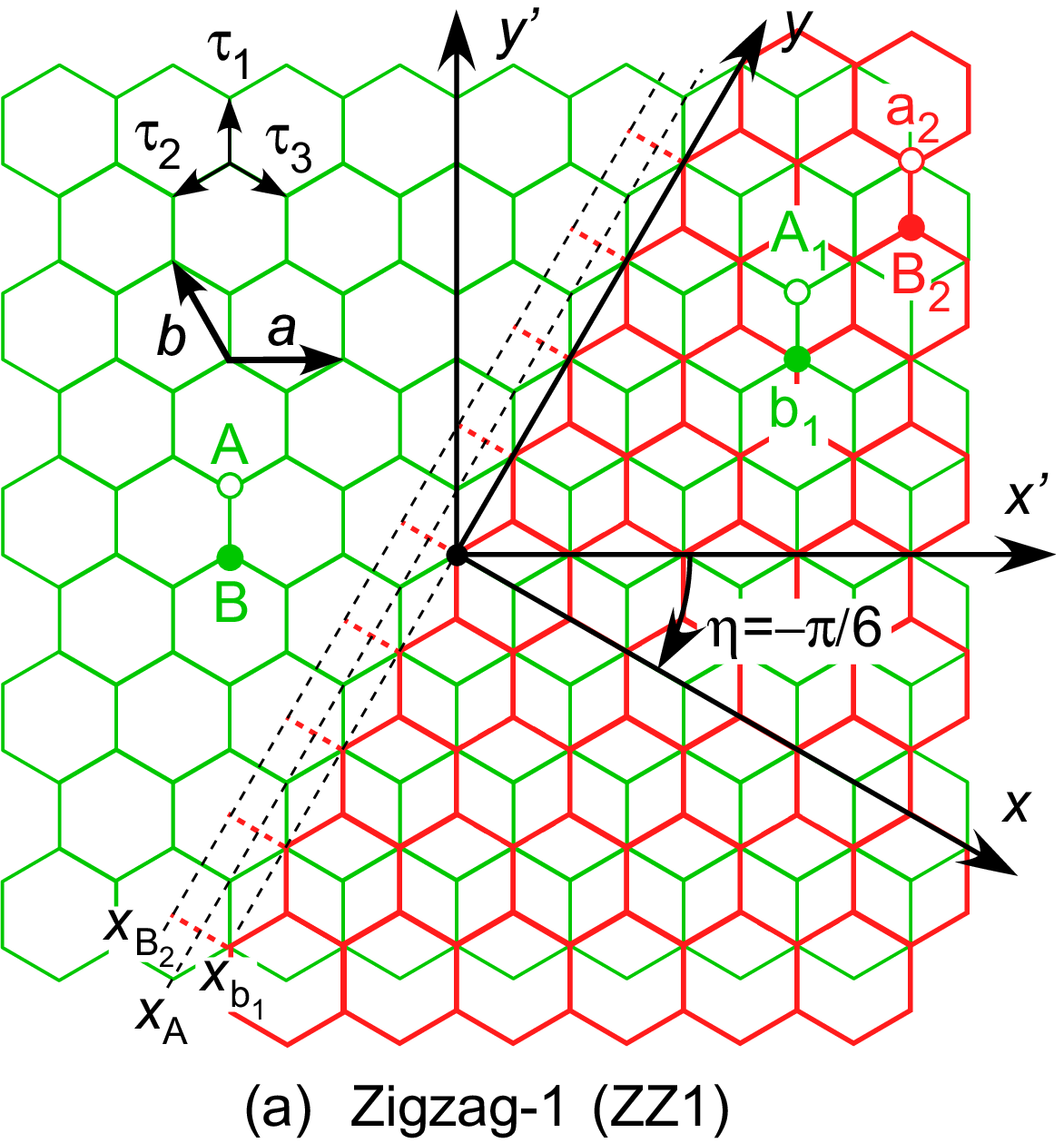} & & \includegraphics[width=6.0cm]{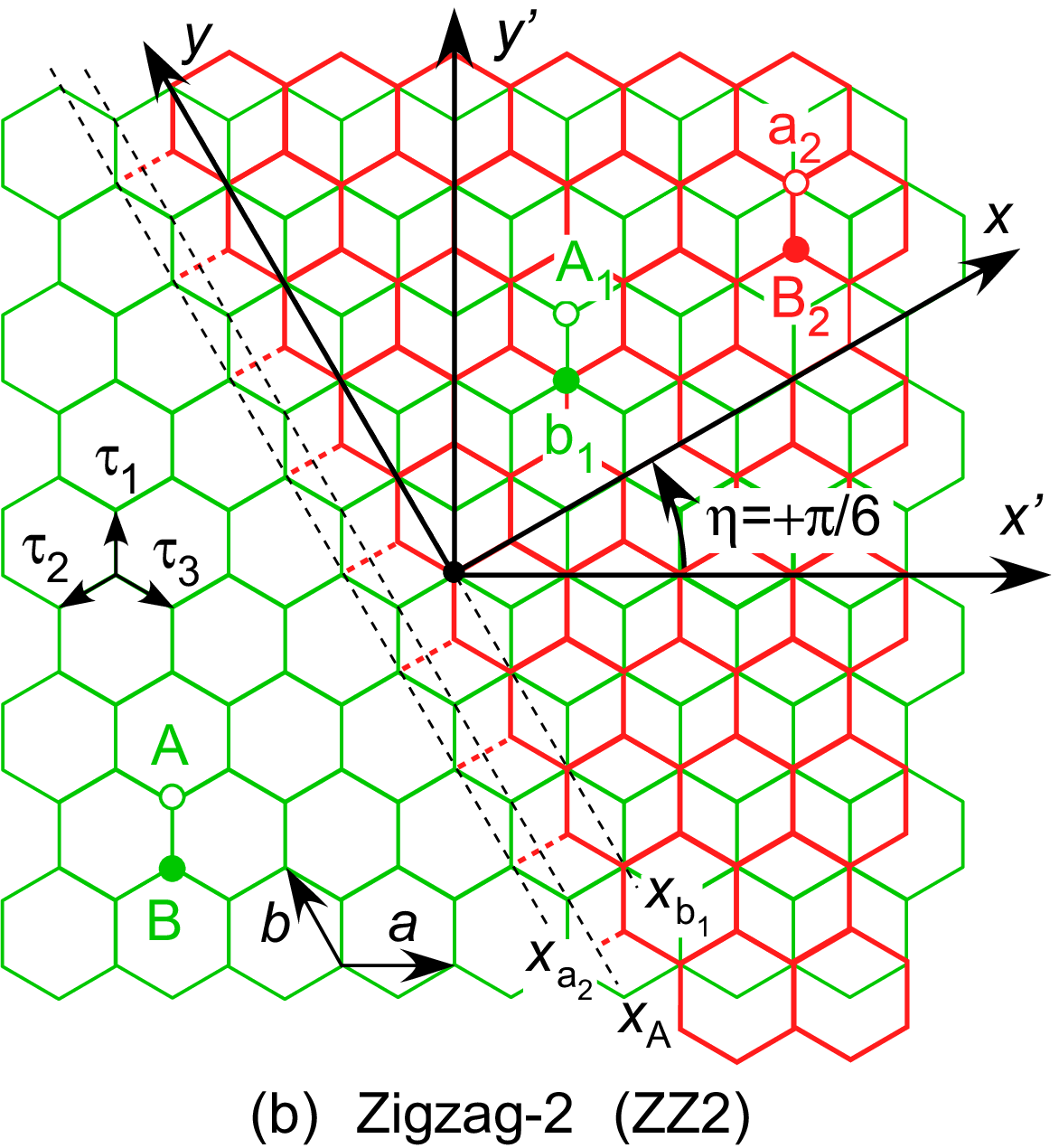} \\
\noalign{\vspace{0.250cm}}
\includegraphics[width=6.0cm]{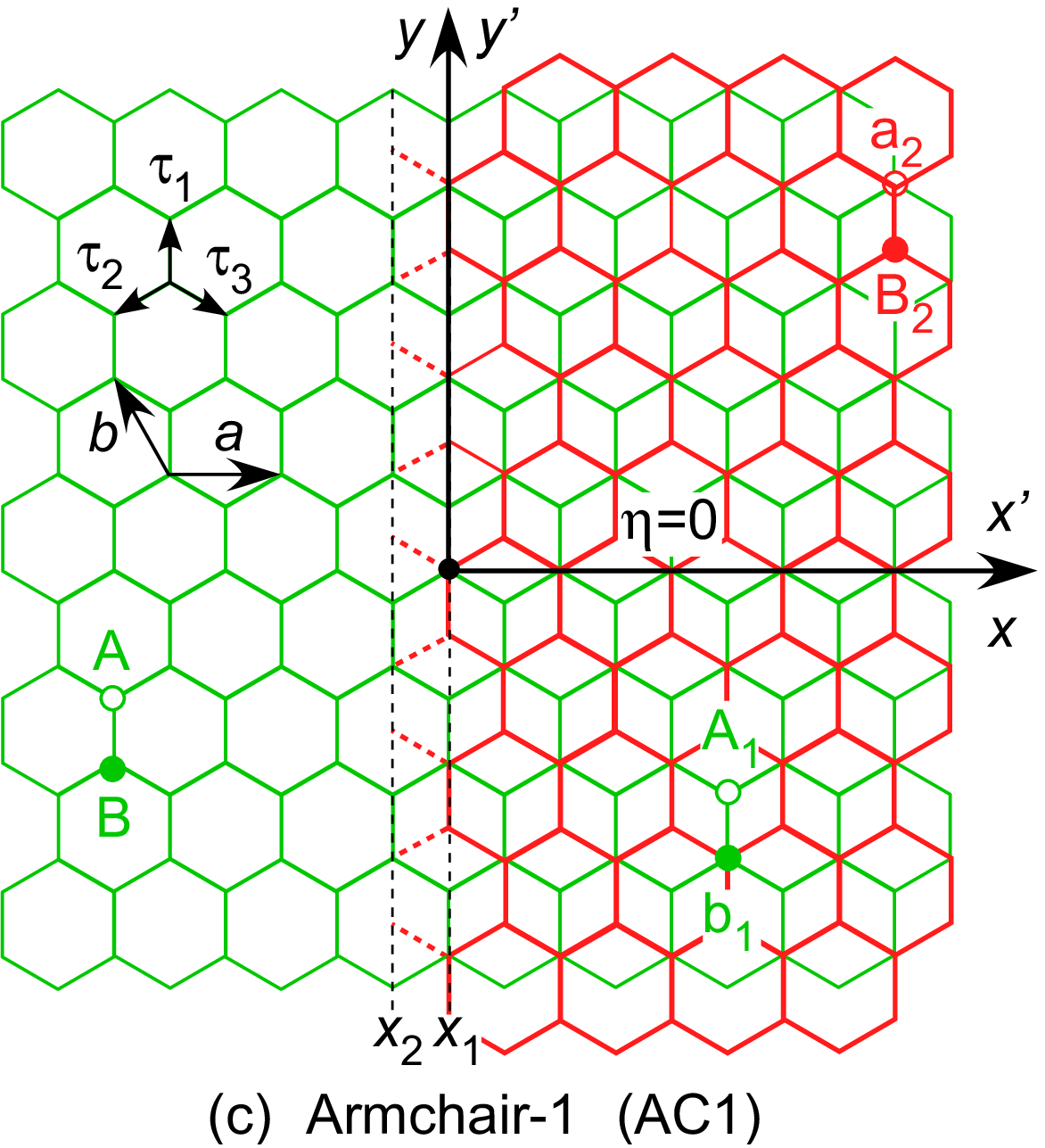} & & \includegraphics[width=6.0cm]{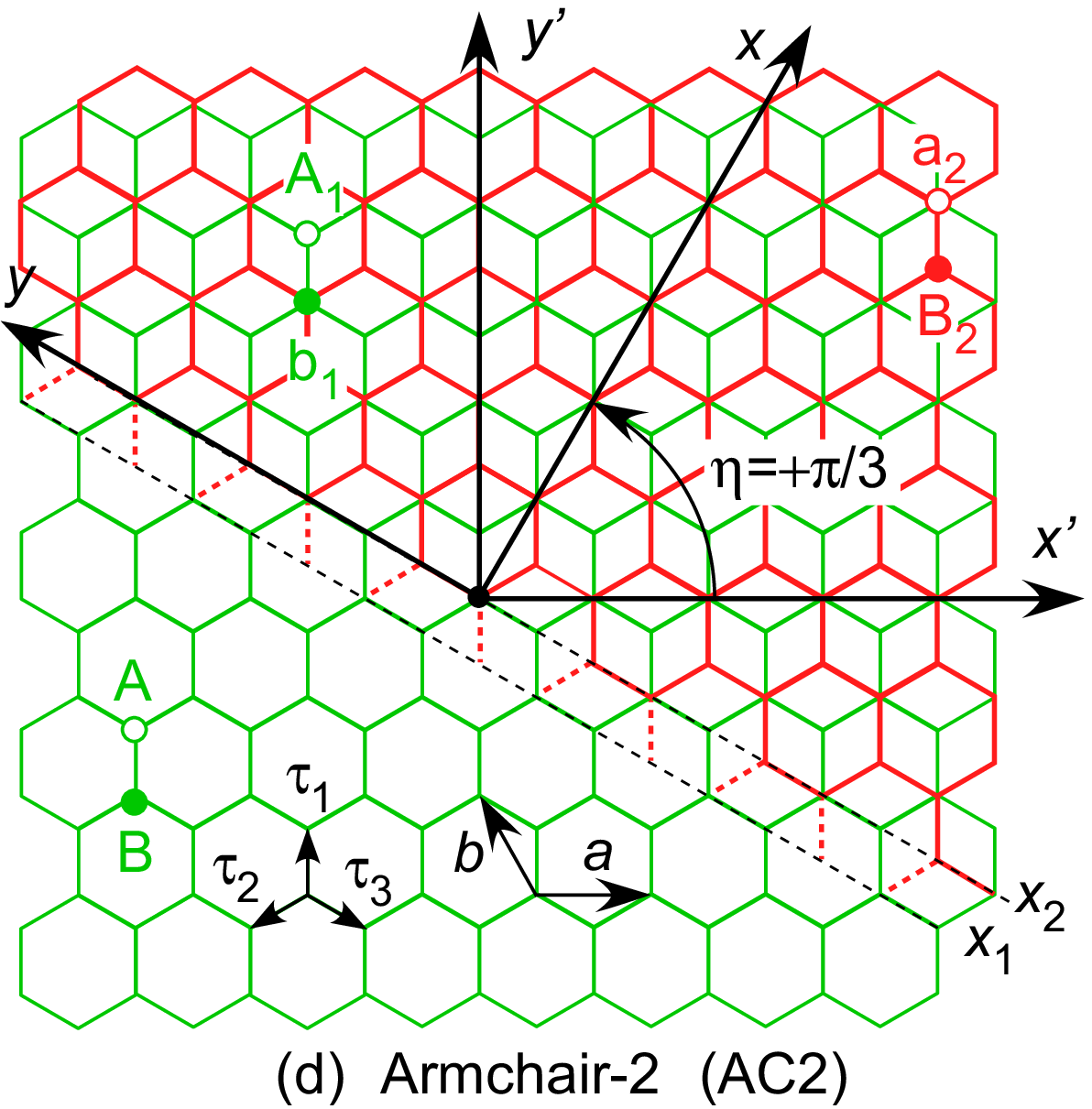}
\end{array}
$$
\caption{(color online)
Atomic structure near boundaries between monolayer and bilayer graphene.
(a) Zigzag boundaries ZZ1 ($\eta=-\pi/6$) and (b) ZZ2 ($\eta=\pi/6$).
(c) Armchair boundaries AC1 ($\eta=0$) and (d) AC2 ($\eta=\pi/3$).
Red (thick) and green (thin) lines represent the top layer with a$_2$ and B$_2$ sites, and bottom layer with A$_1$ and b$_1$ sites, respectively.}
\label{Fig:Stacking_Model}
\end{figure*}
%
\section{Monolayer and Bilayer Graphene}
\subsection{Monolayer graphene}
%
Figure \ref{Fig:Stacking_Model} shows the structure of graphene, two primitive translation vectors ${\bf a}$ and ${\bf b}$, and three vectors $\vec\tau_l$ ($l=1,2,3$) connecting nearest-neighbor atoms.
A unit cell contains two carbon atoms denoted by A and B.
The origin of the coordinates is chosen at a B site, i.e., a B site is given by ${\bf R}_B=n_a{\bf a}+n_b{\bf b}$ and an A site is ${\bf R}_A=n_a{\bf a}+n_b{\bf b}+\vec\tau$ with $n_a$ and $n_b$ being integers and $\vec\tau\!\equiv\!\vec\tau_1=({\bf a}+2{\bf b})/3$.
In the coordinate system $(x',y')$ fixed on the graphene, we have ${\bf a}=a(1,\,0)$, ${\bf b}=a(1/2,\,\sqrt3/2)$, and $\vec\tau=a(0,\,1/\sqrt3)$, where $a=0.246$ nm is the lattice constant.
In the following we start with a tight--binding model with a nearest--neighbor hopping integral $-\gamma_0$.
We consider the coordinates $(x,y)$ rotated around the origin by $\eta$ such that the $y$ axis is always along the boundary of the bilayer graphene.
\par
%
In a monolayer graphene, two bands having approximately a linear dispersion cross at corner K and K' points of the first Brillouin zone.
The wave vectors of the K and K' points are given by ${\bf K}=(2\pi/a)(1/3,\,1/\sqrt3)$ and ${\bf K}'=(2\pi/a)(2/3,\,0)$, respectively.
In a tight-binding model, the wave function is written as
%
\begin{equation}
\psi({\bf r}) = \!\! \sum_{{\bf R}={\bf R}_A} \!\! \psi_A({\bf R}) \phi({\bf r}-{\bf R}) + \!\! \sum_{{\bf R}={\bf R}_B} \!\! \psi_B({\bf R}) \phi({\bf r}-{\bf R}) ,
\end{equation}
%
where $\phi({\bf r})$ denotes a $p_z$ orbital.
The amplitude at atomic sites ${\bf R}={\bf R}_A$ or ${\bf R}_B$ satisfies
%
\begin{equation}
\begin{array}{ll}
\varepsilon \psi_A({\bf R}) & \displaystyle = - \gamma_0 \sum_{l=1,2,3} \psi_B({\bf R} - \vec\tau_l ) , \\
\noalign{\vspace{0.10cm}}
\varepsilon \psi_B({\bf R}) & \displaystyle = - \gamma_0 \sum_{l=1,2,3} \psi_A({\bf R} + \vec\tau_l ) ,
\end{array}
\end{equation}
%
where the overlap integral has been neglected for simplicity.
\par
%
For states in the vicinity of the Fermi level $\varepsilon=0$ of the graphene, the amplitudes are written as 
%
\begin{equation}
\begin{array}{ll}
\psi_{A}({\bf R}) & \! = e^{ i {\bf K}\cdot{\bf R}} F_{A}^{K}({\bf R}) + e^{ i \eta} e^{ i {\bf K}'\cdot{\bf R}} F_{A}^{K'}({\bf R}) , \\
\noalign{\vspace{0.10cm}}
\psi_{B}({\bf R}) & \! = - \omega e^{ i \eta} e^{i {\bf K}\cdot{\bf R}} F_{B}^{K}({\bf R}) + e^{ i {\bf K}'\cdot{\bf R}} F_{B}^{K'}({\bf R}) , 
\end{array}
\label{Eq:Wave_Function_and_Envelopes_in_monolayer_graphene}
\end{equation}
%
in terms of envelope functions $F^K_{A}$, $F^K_{B}$, $F^{K'}_{A}$, and $F^{K'}_{B}$,\cite{Ando_2005a} where $\eta$ is the angle between the $x$ and $x'$ axes as mentioned before and $\omega=e^{2\pi i /3}$.
The envelope functions are assumed to be slowly-varying in the scale of the lattice constant.
\par
%
For the K point, the envelope functions satisfy the Schr\"odinger equation:\cite{Ando_2005a}
%
\begin{equation}
{\cal H}_0 {\bf F}({\bf r}) = \varepsilon {\bf F}({\bf r}) , 
\label{Eq:001}
\end{equation}
%
with
%
\begin{eqnarray}
& {\cal H}_0=\gamma\pmatrix{0 & \hat{k}_{-} \cr \hat{k}_{+} & 0 \cr}, & \label{Eq:002} \\
& {\bf F}^K({\bf r}) = \pmatrix{F^K_{A} ({\bf r})\cr F^K_{B} ({\bf r})\cr }, & \label{Eq:003}
\end{eqnarray}
%
where $\gamma=\sqrt{3}a\gamma_0/2$ is the band parameter, $\hat{k}_{\pm}=\hat{k}_{x}\pm i \hat{k}_{y}$, and $\hat{\bf k}=(\hat{k}_x,\hat{k}_y)=- i \vec\nabla$ is a wave vector operator.
For states with energy $\varepsilon=s\gamma k$ with $s=\pm1$ and $k=\sqrt{k_x^2+k_y^2}$, the wave function is given by
%
\begin{equation}
{\bf F}^K({\bf r}) = \pmatrix{ s k_{-}/k \cr 1 \cr } e^{i {\bf k \cdot r}},
\label{Eq:Envelope_Function:_Monolayer}
\end{equation}
%
apart from a normalization constant.
For the K' point the Schr\"odinger equation is obtained by replacing $\hat k_y$ with $-\hat k_y$ and therefore the wave function by replacing $k_y$ with $-k_y$.
\par
%
\subsection{Bilayer graphene} \label{Ssc:Bilayer_graphene}
%
We consider a bilayer graphene, which is arranged in the AB (Bernal) stacking, as shown in Fig.\ \ref{Fig:Stacking_Model}.
A bottom layer is denoted as 1 and a top layer denoted as 2.
The unit cell contains two carbon atoms denoted by A$_1$ and b$_1$ in layer 1, and a$_2$ and B$_2$ in layer 2.
For the inter-layer coupling, we include coupling $\gamma_1$ between vertically neighboring atoms b$_1$ and a$_2$.
As a result, the states associated with b$_1$ and a$_2$ are pushed away from the Fermi level, which is the reason that they are denoted by lower-case characters.
\par
%
Similar equations of motion can be written down for amplitudes at atomic sites with the use of nearest-neighbor in-plane hopping integral $-\gamma_0$ and inter-layer hopping integral $\gamma_1$.
In terms of slowly-varying envelope functions, the amplitudes are written as
%
\begin{equation}
\begin{array}{rl}
\psi_{A1}({\bf R}) & \! = e^{i {\bf K}\cdot{\bf R}} F_{A1}^{K}({\bf R}) + e^{i \eta} e^{i {\bf K}'\cdot{\bf R}} F_{A1}^{K'}({\bf R}) , \\
\psi_{b1}({\bf R}) & \! = - \omega e^{i \eta} e^{i {\bf K}\cdot{\bf R}} F_{b1}^{K}({\bf R}) + e^{i {\bf K}'\cdot{\bf R}} F_{b1}^{K'}({\bf R}) , \\
\psi_{a2}({\bf R}) & \! = - \omega e^{i \eta}e^{i {\bf K}\cdot{\bf R}} F_{a2}^{K}({\bf R}) + e^{i {\bf K}'\cdot{\bf R}} F_{a2}^{K'}({\bf R}) , \\
\psi_{B2}({\bf R}) & \! = \omega^{-1} e^{2i \eta} e^{i {\bf K}\cdot{\bf R}} F_{B2}^{K}({\bf R}) + e^{-i \eta}e^{i {\bf K}'\cdot{\bf R}} F_{B2}^{K'}({\bf R}) .
\end{array}
\label{Eq:Wave_Function_and_Envelopes}
\end{equation}
%
In the vicinity of the K point, for example, the envelope functions satisfy the Schr\"odinger equation:\cite{Ando_2005a,Koshino_and_Ando_2006a,Snyman_and_Beenakker_2007a}
%
\begin{equation}
{\cal H}_0 {\bf F}^{K}({\bf r}) = \varepsilon {\bf F}^{K}({\bf r}) ,
\label{Eq:Effective_Mass_Equation_in_Bilaer}
\end{equation}
%
with
%
\begin{eqnarray}
& {\cal H}_0 = 
\bordermatrix{ & A1 & b1 & a2 & B2 \cr
& 0 & \!\!\!\!\! \gamma\hat k_{-} \!\!\!\!\! & 0 & 0 \cr
& \gamma\hat k_{+} \!\!\!\!\! & 0 & \gamma_1 & 0 \cr
& 0 & \gamma_1 & 0 & \!\!\!\!\! \gamma\hat k_{-} \cr
& 0 & 0 & \!\!\!\!\! \gamma\hat k_{+} \!\!\!\!\! & 0 \cr }, &
\label{Eq:4_by_4_Hamiltonian_bi-layer_graphene} \\
& {\bf F}^{K}({\bf r}) = \pmatrix{F^K_{A1} ({\bf r})\cr F^K_{b1} ({\bf r})\cr F^K_{a2} ({\bf r})\cr F^K_{B2} ({\bf r})\cr } . &
\end{eqnarray}
%
\par
%
We have two conduction bands ($s=+1)$ and valence bands ($s=-1)$
%
\begin{equation}
\varepsilon_{s,j}(k) = s \bigg[ \pm {1\over 2}\gamma_1+\sqrt{{1\over 4}\gamma_1^2+(\gamma k)^2} \bigg] \quad (j=1,2) ,
\label{Eq:005}
\end{equation}
%
where the lower and upper signs correspond to $j=1$ and $2$, respectively.
In the energy range $-\gamma_1<\varepsilon<+\gamma_1$, in particular, we have a traveling mode corresponding to $\varepsilon_{s,1}(k)$
%
\begin{equation}
{\bf F}^K({\bf r}) = \pmatrix{ \displaystyle - s {k_x - i k_y \over k_x + i k_y } \cr \displaystyle - {\gamma ( k_x - i k_y) \over \gamma_1 + |\varepsilon| } \cr \displaystyle s {\gamma ( k_x - i k_y) \over \gamma_1 + |\varepsilon| } \cr \noalign{\vspace{0.1250cm}} 1 \cr } e^{i {\bf k}\cdot{\bf r} } ,
\label{Eq:Traveling_Mode:Bilayer}
\end{equation}
%
apart from a normalization constant.
We have also evanescent modes decaying or growing exponentially.
The wave function of the decaying mode in the positive $x$ direction, for example, is given by
%
\begin{equation}
{\bf G}^K({\bf r}) = \pmatrix{ \displaystyle s {\kappa_x - k_y \over \kappa_x + k_y } \cr \displaystyle - i { \gamma (\kappa_x - k_y) \over \gamma_1 - |\varepsilon| } \cr \displaystyle - i s {\gamma (\kappa_x - k_y ) \over \gamma_1 - |\varepsilon| } \cr \noalign{\vspace{0.1250cm}} \displaystyle 1 \cr } e^{ - \kappa_x x + i k_y y } ,
\label{Eq:Evanescent_Mode}
\end{equation}
%
with
%
\begin{equation}
\kappa_x = \sqrt{{|\varepsilon|(\gamma_1-|\varepsilon|) \over \gamma^2} + k_y^2 } .
\label{Eq:ImaginaryWaveVector}
\end{equation}
%
\par
%
For the traveling mode, the four-component vector of the wave function for $k_y<0$ is complex conjugate of that for $k_y>0$.
For the evanescent mode, however, the absolute value of the amplitude is quite asymmetric between positive and negative $k_y$.
This asymmetry is the origin of valley polarization of transmitted wave, as will be shown below.
Further, the b$_1$ and a$_2$ components of the evanescent mode diverge at $|\varepsilon|=\gamma_1$, showing that the B$_2$ component vanishes when being properly renormalized.
This is related to the perfect reflection occurring at $|\varepsilon|=\gamma_1$ for some boundaries as will be shown below.
\par
%
In the vicinity of $\varepsilon=0$, i.e., $|\varepsilon|\ll\gamma_1$, the Hamiltonian can be reduced to a (2,2) form with basis set $(A_1, B_2)$ as
%
\begin{equation}
{\cal H}_0=-{\gamma^2\over \gamma_1}\pmatrix{0 & \hat{k}_{-}^2 \cr
 \hat{k}_{+}^2 & 0 \cr} ,
\label{Eq:2_by_2_Hamiltonian_bi-layer_graphene}
\end{equation}
%
where functions $F_{a2}$ and $F_{b1}$ have been eliminated with 
%
\begin{equation}
\begin{array}{ll}
F^K_{b1}({\bf r}) & \displaystyle \approx -{\gamma \over \gamma_1} \hat k_{-} F^K_{B2}({\bf r}) , \\
\noalign{\vspace{0.1250cm}}
F^K_{a2}({\bf r}) & \displaystyle \approx -{\gamma \over \gamma_1} \hat k_{+} F^K_{A1}({\bf r}).
\end{array}
\label{Eq:Decomposition_Approx.}
\end{equation}
%
Corresponding energy eigenvalues are 
%
\begin{equation}
\varepsilon_{s}(k) = s {\gamma^2\over \gamma_1} k^2.
\label{Eq:007}
\end{equation}
%
This effective Hamiltonian describes the second-order process between A$_1$ and B$_2$ {\it via} a$_2$--b$_1$ dimers and reproduces the low-energy part of the dispersion quite well.\cite{McCann_and_Falko_2006a,Koshino_and_Ando_2006a,Katsnelson_2006b,Snyman_and_Beenakker_2007a}
For the evanescent mode given by Eq.\ (\ref{Eq:Evanescent_Mode}) with Eq.\ (\ref{Eq:ImaginaryWaveVector}), we can neglect $|\varepsilon|$ in comparison with $\gamma_1$ in these equations.
\par
%
For the K' point, the Hamiltonian is obtained by the replacements $\hat k_y\rightarrow -\hat k_y$.
Therefore, the wavefunctions are obtained by changing $k_y$ into $-k_y$.
\par
%
\section{Boundary Condition} \label{Sec:Boundary_Condition}
%
Let us consider a boundary between monolayer and bilayer graphene as illustrated in Figs.\ \ref{Fig:Stacking_Model} (a)--(d).
The boundary is straight in the $y$ direction specified by angle $\eta$.
We have zigzag boundaries in both (a) $\eta=-\pi/6$ (ZZ1) and (b) $\eta=\pi/6$ (ZZ2), and armchair boundaries in both (c) $\eta=0$ (AC1) and (d) $\eta=\pi/3$ (AC2).
For these boundaries, the wave functions of both sides can be matched only by those in the vicinity of the K and K' points, given by Eqs.\ (\ref{Eq:Wave_Function_and_Envelopes_in_monolayer_graphene}) and (\ref{Eq:Wave_Function_and_Envelopes}).
In more general cases, boundary conditions involve evanescent states away from the K and K' points, other than those described by Eqs.\ (\ref{Eq:Wave_Function_and_Envelopes_in_monolayer_graphene}) and (\ref{Eq:Wave_Function_and_Envelopes}), and more elaborate treatment is required to derive conditions for the envelope functions.\cite{Akhmerov_and_Beenakker_2008a,Ando_and_Mori_1982a,Ando_et_al_1989a,Ando_and_Akera_1999a}
\par
%
\subsection{Zigzag Boundary: ZZ1} \label{Ssc:Zigzag_Boundary:_ZZ1}
%
First, we consider zigzag boundary ZZ1 with $\eta=-\pi/6$, as shown in Fig.\ \ref{Fig:Stacking_Model} (a).
For A sites on line $x=x_A\equiv-a/(2\sqrt3)$, we have condition:
%
\begin{equation}
\psi_{A}({\bf R}_{A}) = \psi_{A1}({\bf R}_{A}) ,
\quad
{\bf R}_{A} = n({\bf a}+{\bf b}) + \vec\tau_2, 
\label{Eq:Zigzag_Boundary_1-1}
\end{equation}
%
where $\psi_{A1}({\bf R}_A)$ is the wave function extrapolated to ${\bf R}_A$ from the bilayer region.
For b$_1$ sites on line $x=x_{b1}\equiv0$, we have
%
\begin{equation}
\psi_{B}({\bf R}_{b1}) = \psi_{b1}({\bf R}_{b1}) ,
\quad
{\bf R}_{b1} = n ({\bf a} + {\bf b}) ,
\label{Eq:Zigzag_Boundary_1-2}
\end{equation}
%
where $\psi_{B}({\bf R}_{b1})$ is the wave function extrapolated to ${\bf R}_{b1}$ from the monolayer region.
Because of the absence of B$_2$ sites on line $x=x_{B2}\equiv-a/\sqrt3$, we have
%
\begin{equation}
\psi_{B2}({\bf R}_{B2}) = 0 ,
\quad
{\bf R}_{B2} = n ({\bf a} + {\bf b}) - \vec\tau_3 .
\label{Eq:Zigzag_Boundary_1-3}
\end{equation}
%
\par
%
The phase of Bloch functions $e^{i {\bf K}\cdot{\bf R}_A}$ at the K point and $e^{i {\bf K}'\cdot{\bf R}_A}$ at the K' point appearing in Eq.\ (\ref{Eq:Zigzag_Boundary_1-1}) given by Eq.\ (\ref{Eq:Wave_Function_and_Envelopes_in_monolayer_graphene}) rapidly oscillates as a function of $n$ with period of $3$ in a different manner.
Therefore, the condition (\ref{Eq:Zigzag_Boundary_1-1}) is satisfied if and only if the envelope function of each valley is the same along line $x=x_A$, i.e., $F^v_{A1}(x_A,y)=F^v_A(x_A,y)$ for $v=K$ and $K'$.
The same is applicable to Eqs.\ (\ref{Eq:Zigzag_Boundary_1-2}) and (\ref{Eq:Zigzag_Boundary_1-3}), giving $F^v_{b1}(x_{b1},y)=F^v_B(x_{b1},y)$ and $F^v_{B2}(x_{B2},y)=0$ for $v=K$ and $K'$.
Because the envelope functions satisfy first-order differential equations (\ref{Eq:4_by_4_Hamiltonian_bi-layer_graphene}), the boundary conditions are fully specified only by their amplitudes at the boundary.
Therefore, the slight deviation of $x_A$ and $x_{B2}$ from $x=0$ can safely be neglected and the boundary conditions between envelope functions $F^v_{A1}({\bf r}), F^v_{b1}({\bf r})$ and $F^v_{B2}({\bf r})$ in bilayer graphene and $F^v_{A}({\bf r})$ and $F^v_{B}({\bf r})$ in monolayer graphene are written as
%
\begin{equation}
\begin{array}{ll}
F^v_{A1}(0,y) &\! = F^v_A(0,y) , \\
F^v_{b1}(0,y) &\! = F^v_B(0,y) , \\
F^v_{B2}(0,y) &\! = 0 ,
\end{array}
\label{Eq:ZZ1_Envelop}
\end{equation}
%
for $v=K$ and $K'$.
\par
%
The boundary conditions do not cause mixing between the K and K' points, leading to the absence of inter-valley transmission through the boundary.
The transmission of electron wave through the boundary can explicitly be calculated by considering right- and left-going traveling modes (\ref{Eq:Envelope_Function:_Monolayer}) in the monolayer and traveling modes (\ref{Eq:Traveling_Mode:Bilayer}) and an evanescent mode (\ref{Eq:Evanescent_Mode}) decaying in the positive $x$ direction in the bilayer.
Some of the results are presented in Sec.\ V.
\par
%
In order to understand how traveling modes of both sides are connected with each other, we consider the energy region close to the Dirac point $|\varepsilon|\ll\gamma_1$ in the K valley.
Envelope functions in bilayer graphene are composed of traveling waves, to be described by $\tilde{\bf F}^K$, and an evanescent wave ${\bf G}^K$.
The traveling modes in the bilayer side are mainly described by two components $\tilde{F}^K_{A1}$ and $\tilde{F}^K_{B2}$, and other components are eliminated by using Eq.\ (\ref{Eq:Decomposition_Approx.}).
Because the wave vector in the $y$ direction perpendicular to the boundary is conserved, the wave functions are written as $F^K_A({\bf r})=F^K_A(x)e^{i k_yy}$, etc.
After the evanescent mode given by Eq.\ (\ref{Eq:Evanescent_Mode}) being eliminated, we have following boundary conditions for traveling modes
%
\begin{equation}
\pmatrix{F^K_A(0) \cr \noalign{\vspace{0.1250cm}} F^K_B(0) \cr } \approx \pmatrix{ 1 & \displaystyle - s {\kappa_x - k_y \over \kappa_x + k_y } \cr 0 & \displaystyle - {\gamma ( \hat k_x - i \kappa_x) \over \gamma_1 } \cr } \left. \pmatrix{\tilde  F^K_{A1}(x) \cr \noalign{\vspace{0.1250cm}} \tilde{F}^K_{B2}(x) \cr } \right|_{x=0} .
\label{Conditions_for_Major_Components}
\end{equation}
%
Details are discussed in Appendix A.
The boundary conditions for the K' point are obtained by replacing $k_y$ with $-k_y$.
Note that the conditions now include the first derivative of the wave functions in the bilayer side because they satisfy second-order differential equation (\ref{Eq:2_by_2_Hamiltonian_bi-layer_graphene}).
\par
%
In the limit $|\varepsilon|/\gamma_1\rightarrow0$, they are reduced to
%
\begin{equation}
\begin{array}{ll}
F^K_A(0) &\! \displaystyle \approx \tilde{F}^K_{A1}(0) - s {\kappa_x - k_y \over \kappa_x + k_y } \tilde{F}^K_{B2}(0) , \\
F^K_B(0) &\! \approx 0 ,
\end{array}
\label{Eq:ApproximateZZ1}
\end{equation}
%
The amplitude in the bilayer side is asymmetric with respect to the sign of $k_y$, i.e., the direction of the incident wave, and the asymmetry is opposite between the K and K' points.
This means that for waves incident on the interface with oblique angle, transmitted waves have valley polarization.
\par
%
The second condition of Eq.\ (\ref{Eq:ApproximateZZ1}), together with Eq.\ (\ref{Eq:Envelope_Function:_Monolayer}), shows that the reflection coefficient becomes $r_{KK}\approx-1$ and the transmission probability vanishes  when an electron wave is incident from the monolayer side.
On the other hand, the first equation of Eq.\ (\ref{Eq:ApproximateZZ1}) shows that the amplitude of the wave transmitted into the bilayer side is appreciable unless $F^K_A(0)=0$.
These somewhat contradictory conclusions arise from the fact the transmission probability is multiplied by the velocity which is proportional to $k$ in the bilayer side and therefore is much smaller than in the monolayer side.
Some examples of the wave functions will be shown in Fig.\ \ref{Fig:Wave_Function}.
\par
%
\subsection{Zigzag Boundary: ZZ2} \label{Ssc:Zigzag_Boundary:_ZZ2}
%
For the zigzag boundary ZZ2 ($\eta=\pi/6$) illustrated in Fig.\ \ref{Fig:Stacking_Model} (b), boundary conditions become
%
\begin{equation}
\begin{array}{lll}
\psi_A({\bf R}_A) &\! = \psi_{A1}({\bf R}_A) , \enspace & {\bf R}_A=n{\bf b}+\vec\tau_2 , \\
\psi_B({\bf R}_{b1}) &\! = \psi_{b1}({\bf R}_{b1}) , \enspace & {\bf R}_{b1}=n{\bf b} , \\
\psi_{a2}({\bf R}_{a2}) &\! = 0, \enspace & {\bf R}_{a2}=n{\bf b}-{\bf a} ,
\end{array}
\label{Eq:Zigzag_Boundary_2}
\end{equation}
%
giving conditions for the envelope functions
%
\begin{equation}
\begin{array}{ll}
F^v_{A1}(0,y) &\! = F^v_A(0,y) , \\
F^v_{b1}(0,y) &\! = F^v_B(0,y) , \\
F^v_{a2}(0,y) &\! = 0 .
\end{array}
\label{Eq:Boundary_Conditions_ZZ2}
\end{equation}
%
In the vicinity of the Dirac point $|\varepsilon|\ll\gamma_1$, boundary conditions for traveling modes become
%
\begin{equation}
\pmatrix{F^K_A(0) \cr \noalign{\vspace{0.1250cm}} F^K_B(0) \cr } \approx \pmatrix{ \displaystyle 1 \!+\! { i \hat k_x \!-\! k_y \over \kappa_x \!+\! k_y} & 0 \cr \displaystyle \! {s \gamma ( \hat k_x \!+\! i k_y) \over \gamma_1 } & \displaystyle \!\! -\! {\gamma ( \hat k_x \!-\! i k_y) \over \gamma_1 } \! \cr } \!\! \pmatrix{ \tilde{F}^K_{A1}(x) \cr \noalign{\vspace{0.1250cm}} \tilde{F}^K_{B2}(x) \cr } \! \bigg|_{x=0} .
\end{equation}
%
In the limit $|\varepsilon|/\gamma_1\rightarrow0$, they are reduced to
%
\begin{equation}
\begin{array}{ll}
F^K_A(0) &\! \displaystyle \approx \Big( 1 + { i \hat k_x - k_y \over \kappa_x + k_y} \Big) \tilde{F}^K_{A1}(x) \big|_{x=0} , \\
F^K_B(0) &\! \approx 0 .
\end{array}
\label{Eq:ApproximateZZ2}
\end{equation}
%
Essential features of the boundary conditions are the same as in the case of ZZ1.
This fact will be demonstrated by approximate but analytical results in Sec.\ IV and by numerical results in Sec.\ V.
\par
%
\subsection{Armchair Boundary} \label{Ssc:Armchair_Boundary}
%
Next, we consider armchair boundary AC1 ($\eta=0$) shown in Fig.\ \ref{Fig:Stacking_Model} (c).
By a proper extrapolation of the wave functions, we have boundary conditions
%
\begin{equation}
\begin{array}{lll}
\psi_{A}({\bf R}_{A}) &\! = \psi_{A1}({\bf R}_{A}), \enspace &\! {\bf R}_{A}=n({\bf a}+2{\bf b})\!+\!\vec\tau_2 , \\
\psi_{A}({\bf R}_{A1}) &\! = \psi_{A1}({\bf R}_{A1}), \enspace &\! {\bf R}_{A1}=n({\bf a}+2{\bf b})\!+\!\vec\tau_1 , \\
\psi_{B}({\bf R}_{B}) &\! = \psi_{b1}({\bf R}_{B}), \enspace &\! {\bf R}_{B}=n({\bf a}+2{\bf b})+{\bf b} , \\
\psi_{B}({\bf R}_{b1}) &\! = \psi_{b1}({\bf R}_{b1}), \enspace &\! {\bf R}_{b1}=n({\bf a}+2{\bf b}) , \\
\psi_{B2}({\bf R}_{B2}) &\! = 0, \enspace &\! {\bf R}_{B2}=n({\bf a}+2{\bf b})\!-\!\vec\tau_3 , \\
\psi_{a2}({\bf R}_{a2}) &\! = 0, \enspace &\! {\bf R}_{a2}=n({\bf a}+2{\bf b})+{\bf b} ,
\end{array}
\label{Eq:Armchair_Boundary}
\end{equation}
%
where ${\bf R}_{A}$ and ${\bf R}_{B}$ are on line $x=x_2\equiv-a/2$, ${\bf R}_{A1}$ and ${\bf R}_{b1}$ are on $x=x_1\equiv0$, and ${\bf R}_{B2}$ and ${\bf R}_{a2}$ are on $x=x_2$.
Because ${\bf K}\cdot({\bf a}+2{\bf b})={\bf K}'\cdot({\bf a}+2{\bf b})=0$ (mod $2\pi$), the Bloch functions remain constant on lines $x=x_2$ and $x=x_1$.
Thus, we have from the first and second conditions of Eq.\ (\ref{Eq:Armchair_Boundary})
%
\begin{equation}
\begin{array}{rl}
{\bf F}^{K}_{A1}({\bf r}) + {\bf F}^{K'}_{A1}({\bf r}) &\! = {\bf F}^{K}_{A}({\bf r}) + {\bf F}^{K'}_{A}({\bf r}) \big|_{x=x_2} , \\
\noalign{\vspace{0.1250cm}}
\omega {\bf F}^{K}_{A1}({\bf r}) + {\bf F}^{K'}_{A1}({\bf r}) &\! = \omega {\bf F}^{K}_{A}({\bf r}) + {\bf F}^{K'}_{A}({\bf r}) \big|_{x=x_1} ,
\end{array}
\end{equation}
%
respectively.
Note that the slight deviation of $x_1$ and $x_2$ from $x=0$ can safely be neglected from the same argument for zigzag boundary.
Because envelope functions are slowly varying in the scale of a lattice constant, both conditions are satisfied, if and only if they are the same within each valley.
Exactly the same argument is applicable to the third and fourth conditions.
For the fifth and sixth conditions of (\ref{Eq:Armchair_Boundary}), we use $e^{i({\bf K}'-{\bf K}) \cdot {\bf R}_{B2}}=\omega^{-1}$ and $e^{i ({\bf K}'-{\bf K}) \cdot {\bf R}_{a2}}=\omega$.
Then, the boundary conditions for the envelope functions are summarized as
%
\begin{equation}
\begin{array}{rl}
F^v_{A1}(0,y) &\! = F^v_A(0,y) , \\
F^v_{b1}(0,y) &\! = F^v_B(0,y) , \\
F^K_{a2}(0,y) - F^{K'}_{a2}(0,y) &\! = 0 , \\
F^K_{B2}(0,y) + F^{K'}_{B2}(0,y) &\! = 0 .
\end{array}
\label{Eq:BoundaryConditions_AC1}
\end{equation}
%
\par
%
Armchair boundary AC2 of $\eta=\pi/3$ is illustrated in Fig.\ \ref{Fig:Stacking_Model} (d).
In a similar manner, the boundary conditions are obtained as
%
\begin{equation}
\begin{array}{rl}
F^v_{A1}(0,y) &\! = F^v_A(0,y) , \\
F^v_{b1}(0,y) &\! = F^v_B(0,y) , \\
\omega F^K_{a2}(0,y) + F^{K'}_{a2}(0,y) &\! = 0 , \\
\omega F^K_{B2}(0,y) - F^{K'}_{B2}(0,y) &\! = 0 .
\end{array}
\label{Eq:BoundaryConditions_AC2}
\end{equation}
%
These conditions are converted into those of AC1 (\ref{Eq:BoundaryConditions_AC1}) by changing the relative phases of the envelope functions for the K and K' points.
Therefore, there is no difference between transmission probabilities, etc.\ of AC1 and AC2 within the present {\bf k}$\cdot${\bf p} scheme, although actual wave functions $\psi_A({\bf R})$, etc.\ may be different.
\par
%
Inter-valley mixing occurs at the armchair boundary in contrast to the zigzag boundaries.
Effective boundary conditions in the vicinity of the Dirac point, $|\varepsilon|\ll\gamma_1$, can be derived in a manner similar to those for the zigzag boundaries and the results are presented in Appendix \ref{Sec:Armchair_Boundary}.
There, we show that the conditions are essentially similar except for the presence of small inter-valley mixing.
In fact, in the limit of $k\rightarrow 0$, an injected wave is perfectly reflected within each valley, i.e., $r_{KK}=-1$ and $r_{K'K}=0$ for wave incident in the K valley and the transmission increases with energy as for the zigzag boundaries.
\par
%
\begin{table}[!t]
\begin{center}
\begin{tabular}{lccccl}
\hline
\noalign{\vspace{0.050cm}}
\hline
 & \multicolumn{2}{c}{K} & \multicolumn{2}{c}{K'} \\ 
\noalign{\vspace{-0.10cm}}
 & \enspace $k_y>0 \,$ & $\, k_y<0$ \enspace & \enspace $k_y>0 \,$ & $\, k_y<0$ \enspace & Amplitude \\
\hline
 ZZ1 & 0 & 1 (1) & 1 (1) & 0 & A$_1$, a$_2$ \\
\hline
 ZZ2 & 1 & 0 & 0 & 1 & B$_2$ \\
\hline
 AC1/AC2\hspace{-0.250cm} & \multicolumn{2}{c}{0} & \multicolumn{2}{c}{0} & \\
\hline
\end{tabular}
\caption{The number of $\varepsilon=0$ edge states present in the bilayer graphene localized at the boundaries of monolayer and bilayer graphenes.
The carbon sites where the wave function has nonzero amplitude are shown in the rightmost column.
The number in parenthesis denotes that of perfectly reflecting states present at $|\varepsilon|=\gamma_1$.}
\label{Tab:Number_of_Edge_States}
\end{center}
\vspace{-0.50cm}
\end{table}
%
\subsection{Edge States and Perfectly Reflecting States}
%
As in monolayer and bilayer graphenes, there exists an edge state at $\varepsilon=0$ with amplitude only in the bilayer region localized at the boundary for zigzag boundaries (ZZ1 and ZZ2) and no edge state for armchair boundaries as is shown in Table \ref{Tab:Number_of_Edge_States}.
The details on the derivation are discussed in Appendix \ref{Sec:Edge_States}.
These edge states do not play important roles in the transmission through the boundaries because the transmission is possible only away from $\varepsilon=0$.
In Appendix \ref{Sec:Edge_States}, further, we show that at $|\varepsilon|=\gamma_1$ we have perfect reflection in the region $k_y<0$ at the K point and $k_y>0$ at the K' point only for boundary ZZ1.
These states are also included in Table \ref{Tab:Number_of_Edge_States}.
This special feature of ZZ1 clearly appears in numerical results presented in Sec.\ V.
\par
%
\section{Valley Polarization} \label{Sec:Valley_Polarization}
%
We consider electron transmission between a monolayer and bilayer graphene with same electron concentration.
This is realized when the electron density is changed by a gate voltage.
In the presence of electric field due to gate, the symmetry between the top and bottom layers of a bilayer graphene is broken and a small band gap can open.\cite{McCann_2006a,Min_et_al_2007a,Ando_and_Koshino_2009a,Ando_and_Koshino_2009b}
This small gap will be completely neglected in the following, because we are interested in the essential feature of the transmission property.
Besides, the Fermi level always lies away from the gap and the asymmetry can be controlled by the field due to an extra gate.
The electron density higher than $\gamma k/\gamma_1>\sqrt2$ can experimentally be achieved by various methods.\cite{Das_et_al_2009a,Miyazaki_et_al_2010a}
\par
%
Electron wave with wave vector ${\bf k}$ and positive group velocity in the $x$ direction at Fermi energy $\varepsilon_F$ is injected from the K valley in the monolayer side at the Fermi level.
For incident angle $\theta$ ($-\pi/2<\theta<\pi/2$), we have $k_x=s k\cos\theta$ and $k_y=s k\sin\theta$ with $k=|{\bf k}|$ for the incident wave.
The wave is reflected in the direction  $\pi-\theta$.
\par
\begin{figure}[!t]
\begin{center}
\includegraphics[width=7.50cm]{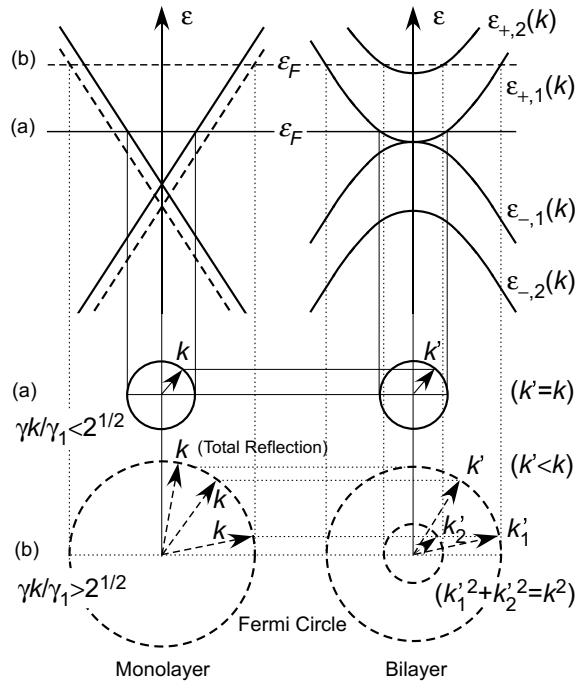}
\caption{Some examples of the alignment of energy bands and corresponding Fermi circles in monolayer and bilayer graphene under the condition of equal electron concentration.}
\label{Fig:Band_Alignment_and_Fermi_Lines}
\end{center}
\vspace{-0.50cm}
\end{figure}

%
When $\gamma k/\gamma_1<\sqrt2$, only a single conduction band is occupied in the bilayer.
In this case, the wave transmitted into the bilayer has the same wave vector ${\bf k}$, i.e., there is no refraction.
When $\gamma k/\gamma_1>\sqrt2$, two bands are occupied by electrons in the bilayer, giving rise to two Fermi circles.
In this case, the number of transmitted waves changes from two to one with the increase of $\theta$ and the total reflection occurs for sufficiently large $\theta$.
This is illustrated in Fig.\ \ref{Fig:Band_Alignment_and_Fermi_Lines}.
\par
%
%
In the energy region close to the Dirac point $\gamma k/\gamma_1\rightarrow0$, a simple expression can be obtained for the amplitude $f$ of the transmitted wave for incident wave given by Eq.\ (\ref{Eq:Envelope_Function:_Monolayer}).
The details on the derivation are discussed in Appendix A.
The result is
%
\begin{equation}
f \approx f_0 \equiv {2 s \cos\theta\over  -e^{-2 i \theta} - (\sin\theta -s\sqrt{1+\sin^2\theta})^2 } .
\label{Eq:Ratio_of_wave:_ZZ}
\end{equation}
%
Because the velocity is $\gamma/\hbar$ in the monolayer and $2\gamma^2 k/\gamma_1\hbar$ in the bilayer, the transmission probability is proportional to $k|f|^2$.
Therefore, it vanishes for $k\!=\!0$ in agreement with $r_{KK}=-1$ as discussed in the previous section and increases in proportion to $k$.
Further, it takes a maximum at $\theta=s\theta_0$, with
%
\begin{equation}
\theta_0 = \sin^{-1}{1\over\sqrt{3}} \approx 0.196 \pi.
\label{Eq:Maximum}
\end{equation}
%
\par
%
For the K' point the amplitude is obtained by replacing $\theta$ with $-\theta$.
The valley polarization\cite{Rycerz_et_al_2007a} of the transmitted wave becomes
%
\begin{equation}
P = {T_K-T_{K'} \over T_K+T_{K'}} = s {2\sin\theta\sqrt{1+\sin^2\theta}\over 1+2\sin^2\theta} ,
\label{Eq:Valley_polarization}
\end{equation}
%
where $T_K$ and $T_{K'}$ are transmission probability into K and K' valley, respectively.
The valley polarization increases with incident angle $\theta$, up to $P=\pm 2\sqrt{2}/3\approx \pm 0.94$ at $\theta=\pm \pi/2$.
\par
%
\begin{figure*}[t]
\begin{center}
\hbox{\includegraphics[width=6.60cm]{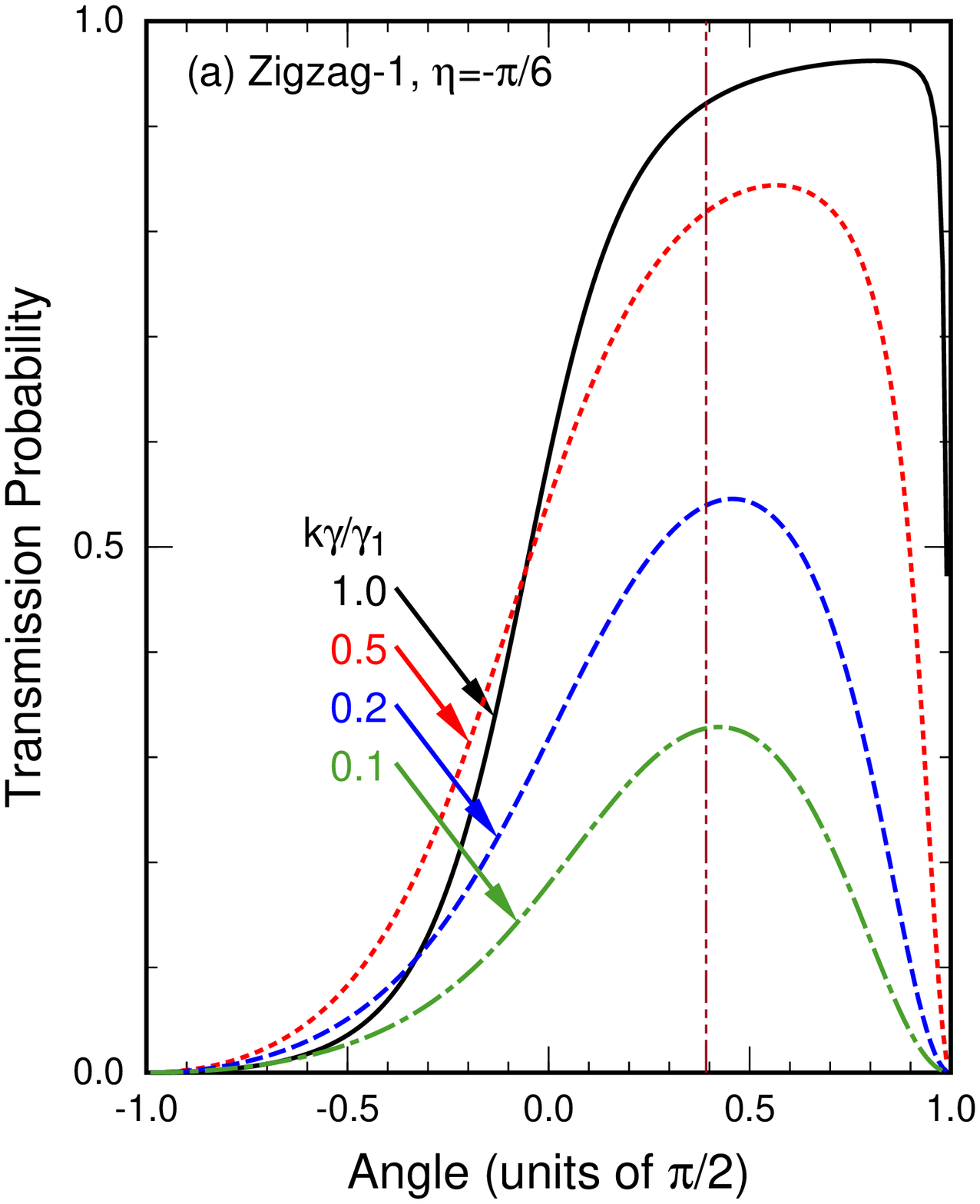}
\hspace{-1.1250cm}
\includegraphics[width=6.60cm]{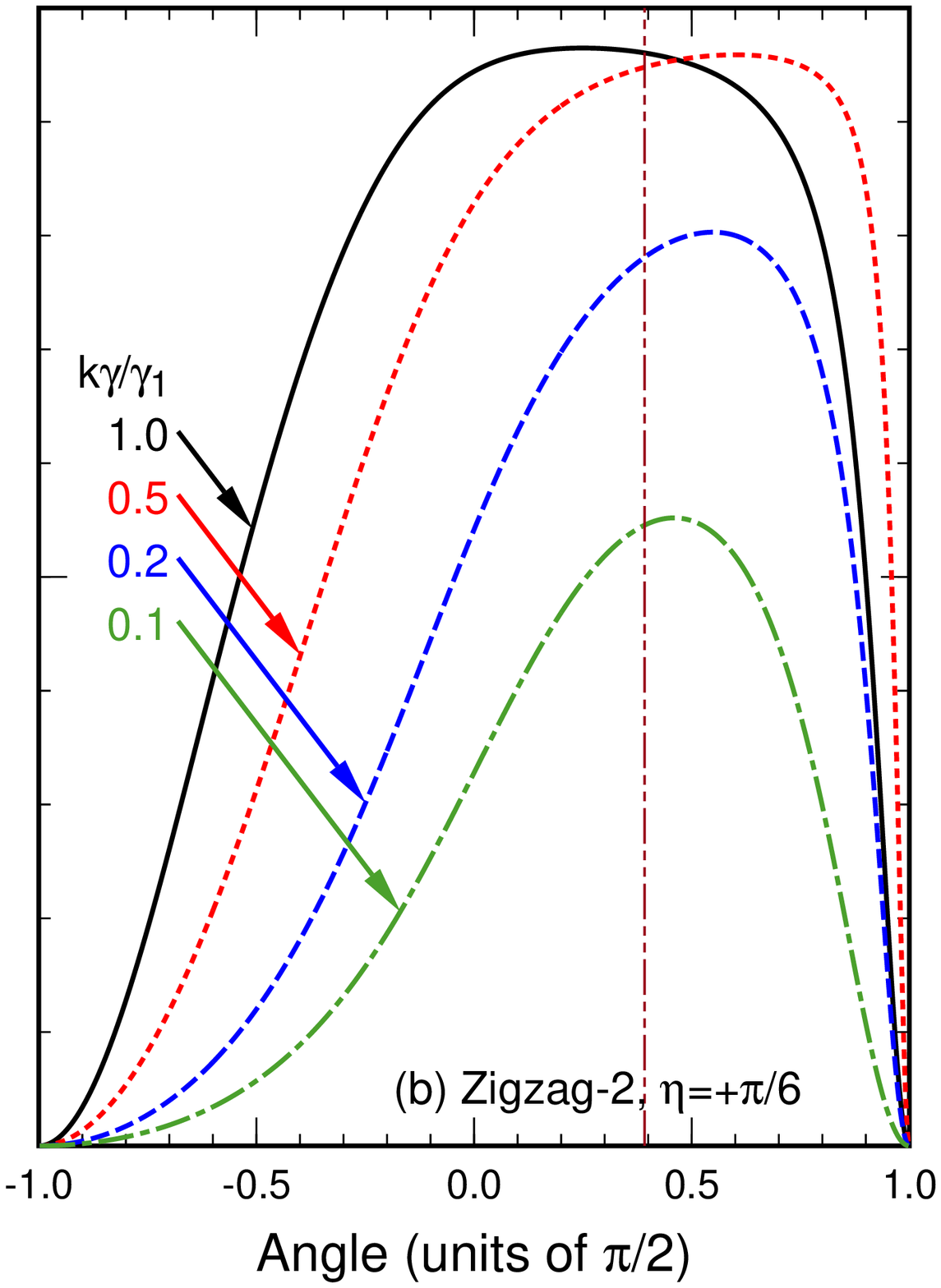}
\hspace{-1.1250cm}
\includegraphics[width=6.60cm]{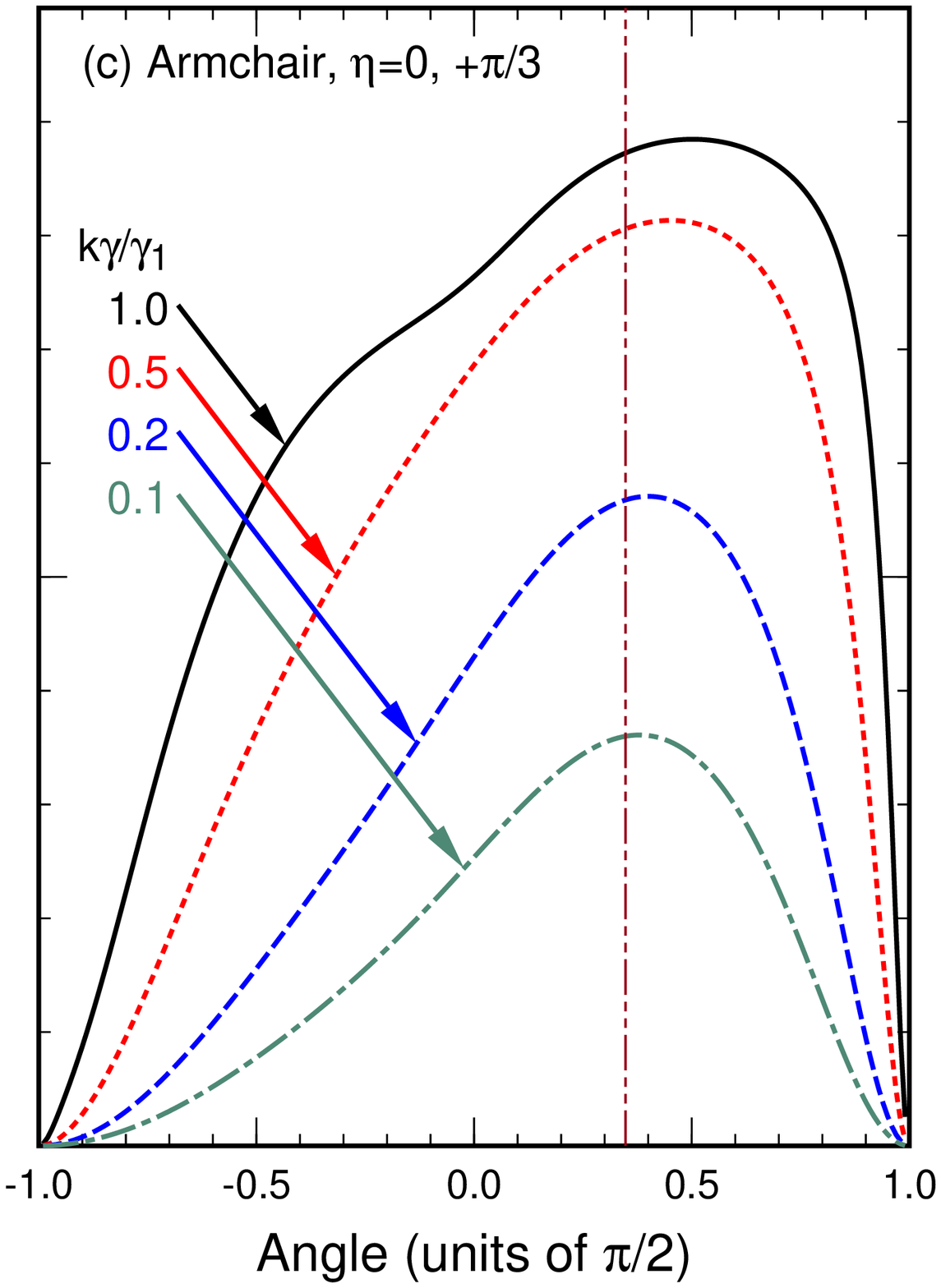}}
\caption{(color online)
Calculated transmission probabilities of the K valley as a function of incident angle $\theta$ for several charge densities specified by $k$ in the monolayer graphene.
(a) A zigzag boundary (ZZ1) with $\eta=-\pi/6$.
(b) A zigzag boundary (ZZ2) with $\eta=+\pi/6$.
(c) An armchair boundary (AC1 and AC2) with $\eta=0$ and $+\pi/3$.
The vertical dot-dot-dashed line shows the maximum-angle in the limit $k\rightarrow0$.}
\label{Fig:Transmission_vs_Incident_Angle}
\end{center}
\vspace{-0.50cm}
\end{figure*}
%
For a ZZ2 boundary, from the first equation of (\ref{Eq:ApproximateZZ2}), the amplitude is calculated as
%
\begin{equation}
f = {2 s \cos\theta \over -e^{-2 i \theta} + i e^{- i \theta}(\sin\theta -s \sqrt{1+\sin^2\theta}) } .
\end{equation}
%
We have
%
\begin{equation}
f = \sqrt{2} f_0 e^{i(\theta+\alpha)} , \quad
\alpha =-s \tan^{-1}{{\sqrt{1+\sin^2\theta} \over \cos\theta}} ,
\label{Eq:Zigzag 2 amplitude} \\
\end{equation}
%
where $f_0$ is defined in Eq.\ (\ref{Eq:Ratio_of_wave:_ZZ}) for ZZ1 boundary.
Therefore, a maximum transmission also occurs at $\theta=s\theta_0$ for the K point and $-s\theta_0$ for the K' point.
For armchair boundaries, the analytic expression of the amplitude is presented in Appendix \ref{Sec:Armchair_Boundary}.
It shows that the inter-valley transmission probability between K and K' is 1/5 of the intra-valley transmission for perpendicularly incident wave ($\theta=0$) near the Dirac point and that maximum transmission occurs at $\theta\approx s \times 0.179\pi$ for the K point and $\theta\approx- s \times 0.179\pi$ for the K' point.
\par
%
The valley polarization completely disappears when two traveling waves are involved in the transmission in the bilayer, i.e., for small incident angles in the case $\gamma k/\gamma_1>\sqrt2$.
In this case the wave functions for $-k_y$ are simply obtained by taking complex conjugate of those for $k_y$ in both monolayer and bilayer graphenes and therefore the reflection and transmission coefficients for $-\theta$ are related to those for $+\theta$ through complex conjugate.
Consequently, the transmission and reflection probabilities become symmetric about $\theta=0$, as will be demonstrated in the next section.
Asymmetry reappears at large incident angle for which transmission into a single traveling wave is allowed.
\par
%
\section{Numerical Results} \label{Sec:Numerical_Results}
%
Figure \ref{Fig:Transmission_vs_Incident_Angle} shows some examples of calculated transmission probability as a function of incident angle for (a) zigzag boundary ZZ1 with $\eta=-\pi/6$, (b) zigzag ZZ2 with $\eta=+\pi/6$, and (c) armchair (AC1 and AC2) with $\eta=0$ and $+\pi/3$.
The electron density is specified by $k$ corresponding to the Fermi energy in the monolayer and the results in the low-density regime $\gamma k/\gamma_1<\sqrt2$ are shown.
The transmission probability varies strongly as a function of the incident angle and its maximum appears at an angle deviating from the vertical direction.
This asymmetry is opposite between the K and K' points, showing that strong valley polarization can be induced across the interface of monolayer and bilayer graphenes.
Except in the high-concentration region $\gamma k/\gamma_1\sim\sqrt2$, the valley polarization is similar for different boundaries.
\par
%
\begin{figure*}[!t]
\begin{center}
\hbox{\includegraphics[width=6.60cm]{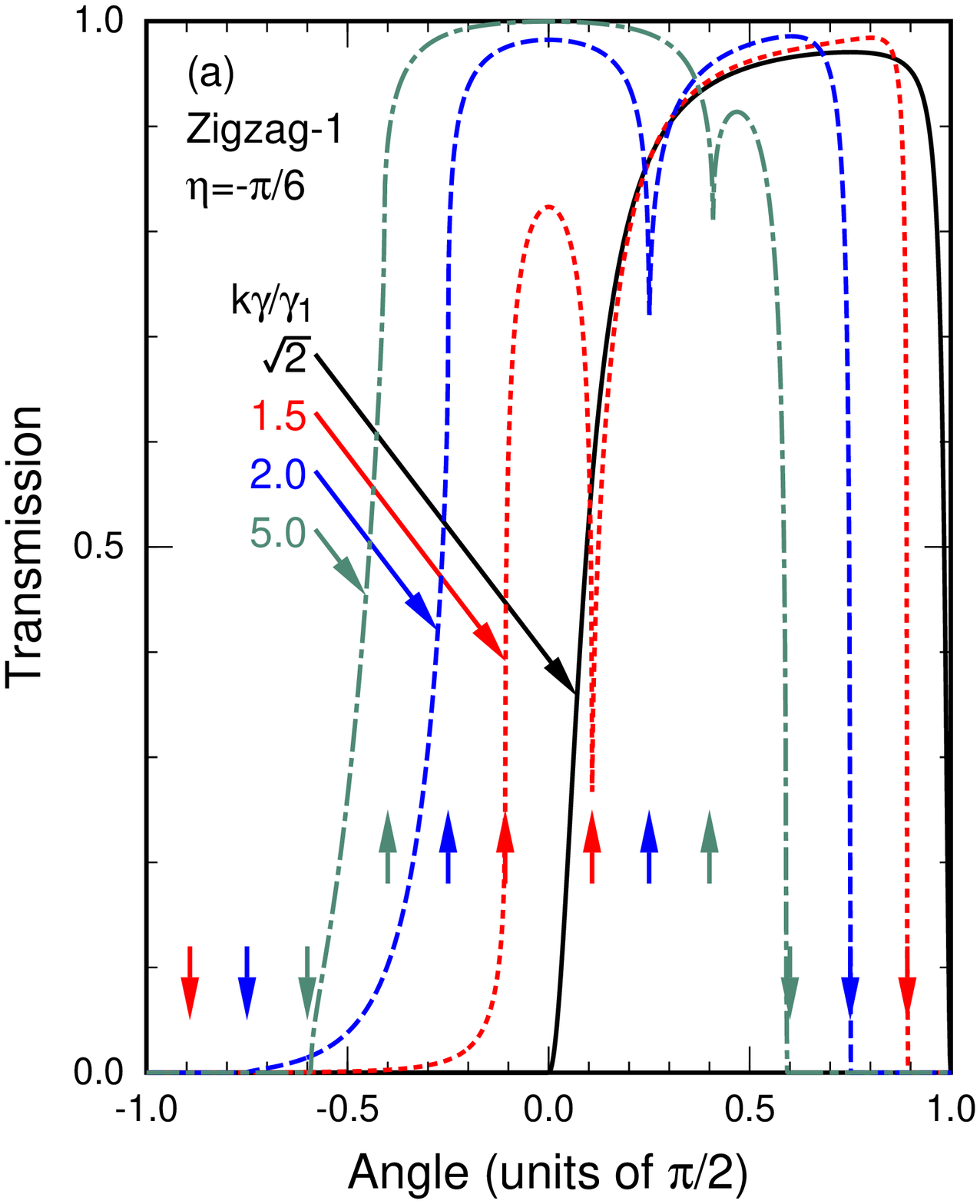}
\hspace{-1.1250cm}
\includegraphics[width=6.60cm]{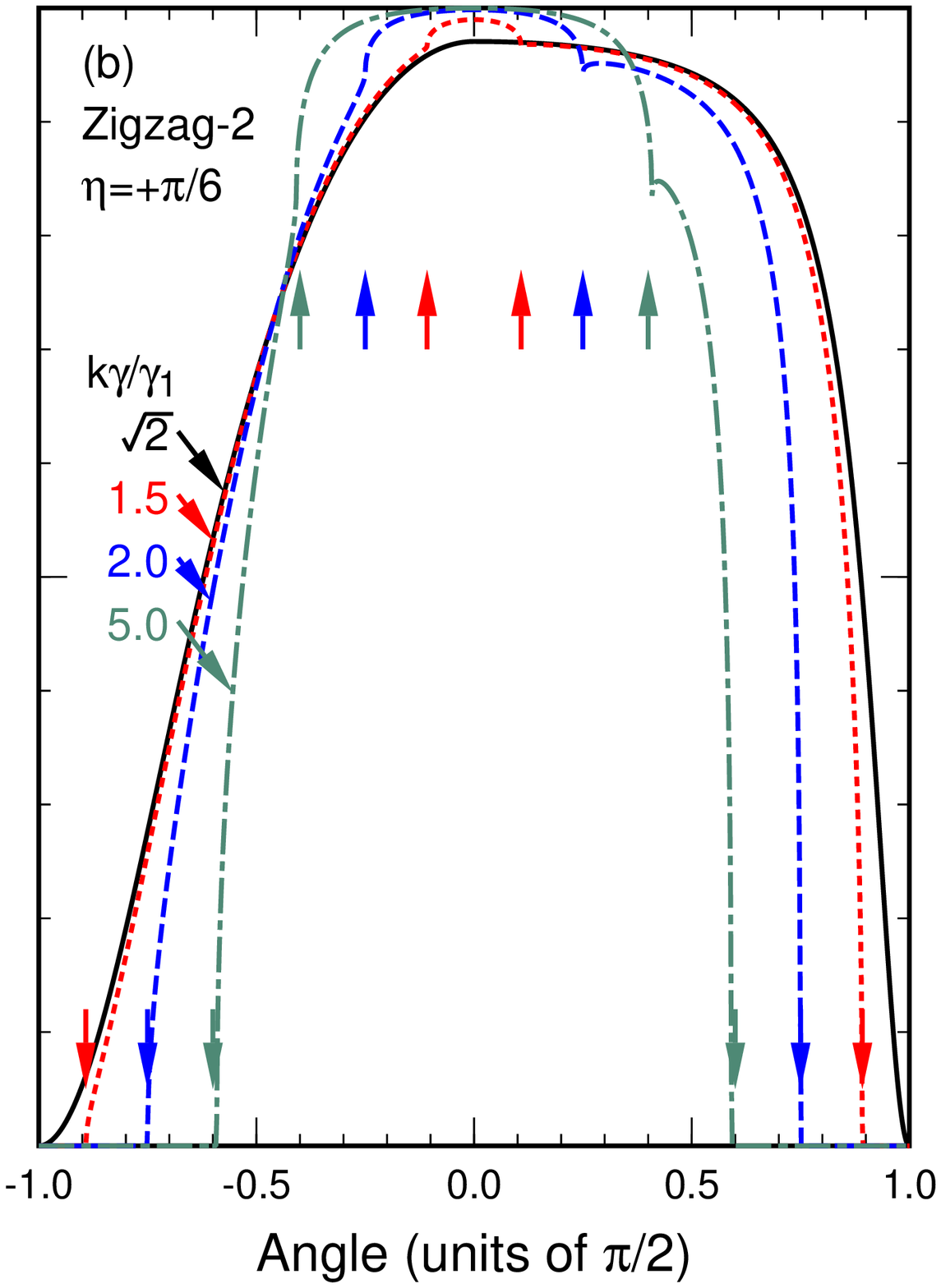}
\hspace{-1.1250cm}
\includegraphics[width=6.60cm]{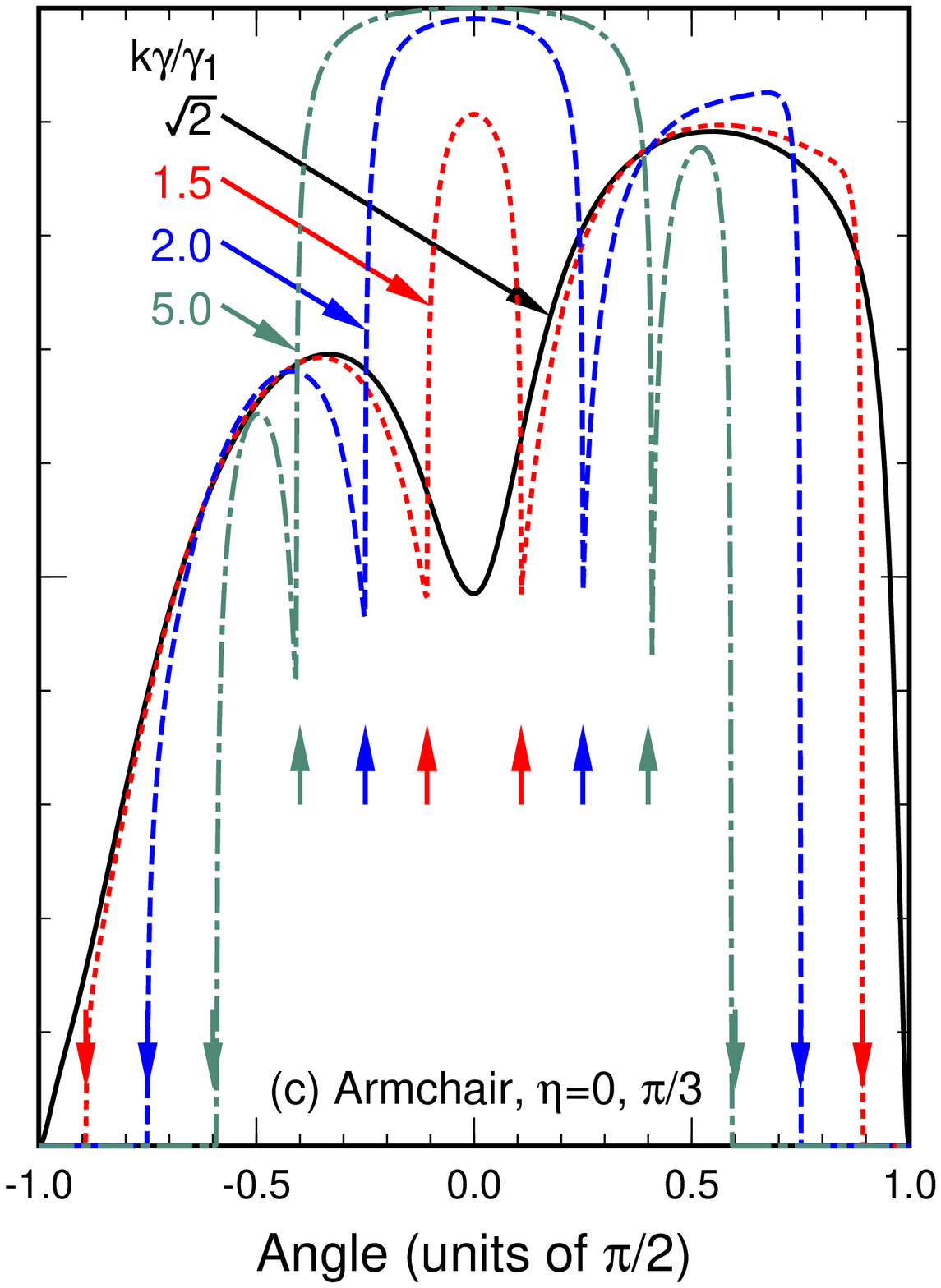}}
\vspace{-3.2500cm}
\hbox{\hspace{7.9500cm}\includegraphics[width=2.90cm]{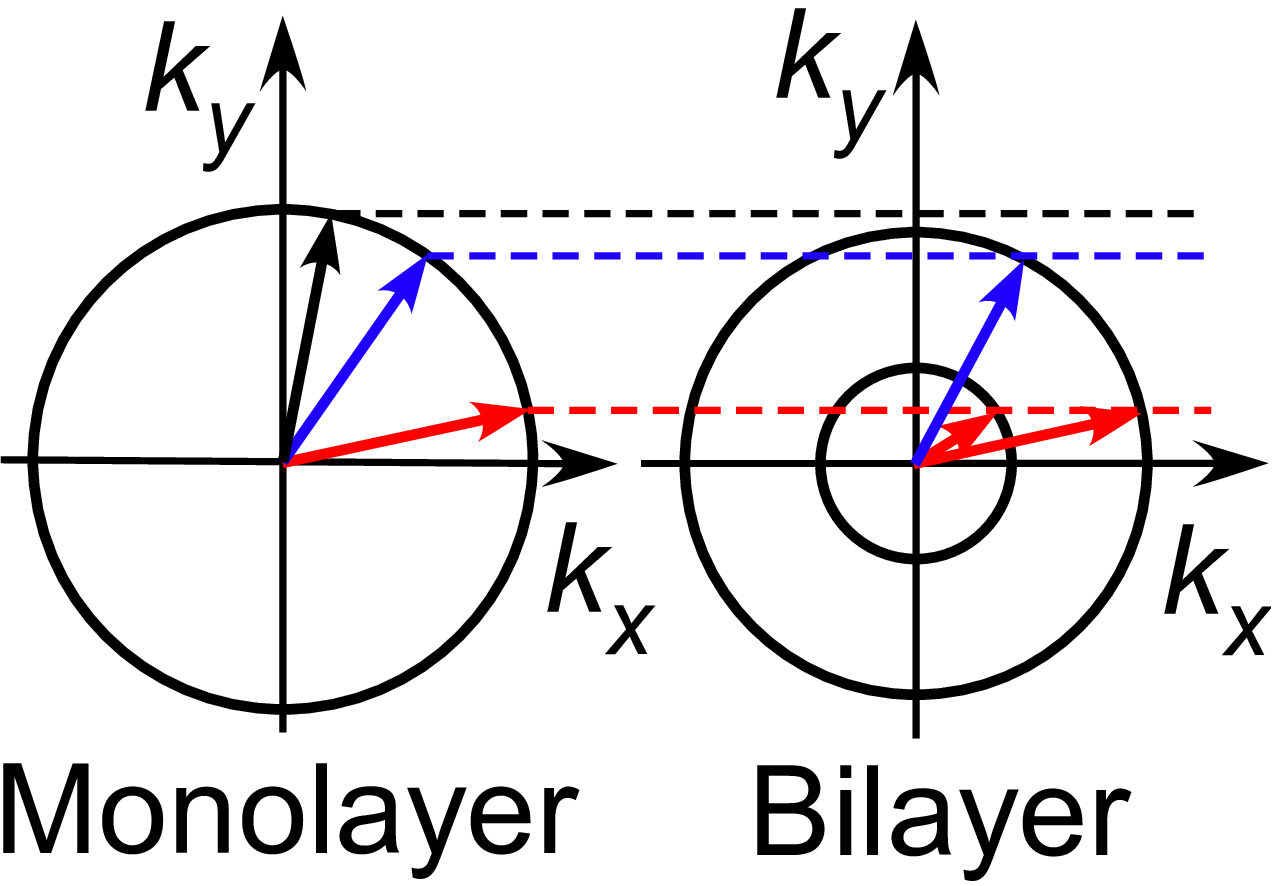}}
\vspace{1.2500cm}
\caption{(color online)
Calculated transmission probabilities for $\gamma k/\gamma_1\geq \sqrt2$, corresponding to Fig.\ \ref{Fig:Transmission_vs_Incident_Angle}.
Fermi lines in monolayer and bilayer graphene are depicted in the inset.
Two transmitted waves are present in the bilayer in the region between two upward arrows and no transmission is allowed outside the region denoted by downward arrows.}
\label{Fig:Transmission_vs_Incident_Angle2}
\end{center}
\vspace{-0.50cm}
\end{figure*}
%
\begin{figure}[!t]
\begin{center}
\begin{center}
\includegraphics[width=7.250cm]{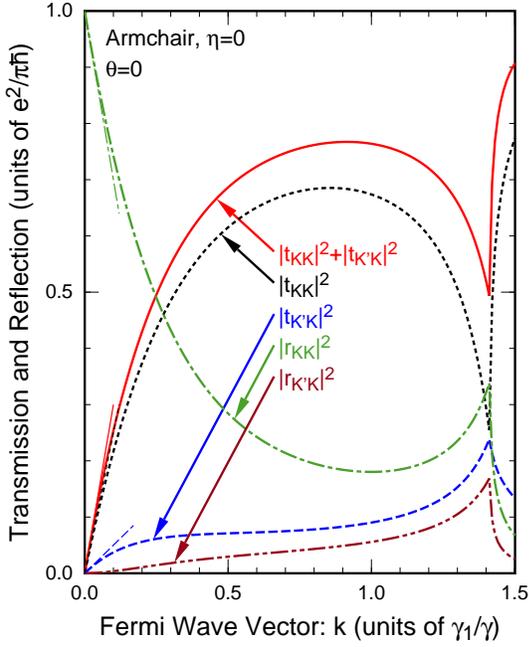}
\caption{(color online)
Calculated transmission and reflection probabilities through armchair boundary AC1 as a function of the Fermi wave length in the monolayer graphene.
Thin lines show analytic results valid for $\gamma k/\gamma_1\!\ll\!1$.
}
\label{Fig:Charge_Density:_Armchair}
\end{center}
\end{center}
\vspace{-0.50cm}
\end{figure}
%
Figure \ref{Fig:Transmission_vs_Incident_Angle2} shows the total transmission probability in the high-density region $\gamma k/\gamma_1\geq\sqrt2$.
The transmission probability depends strongly on boundaries.
In fact, at the bottom of the first excited conduction band, i.e., $k \gamma/\gamma_1=\sqrt{2}$, it completely vanishes in the region $\theta\leq 0$ for ZZ1, but not for ZZ2 and armchair boundaries.
This vanishing transmission at $\varepsilon\!=\!\gamma_1$ for boundary ZZ1 is closely related to the presence of a perfectly reflecting state in the region $k_y<0$ and $k_y>0$ for the K and K' point, respectively, as discussed in Sec.\ III.D and Appendix \ref{Sec:Edge_States}.
\par
%
This can also be understood directly from the behavior of the evanescent mode given by (\ref{Eq:Evanescent_Mode}) and the boundary condition.
At $|\varepsilon|\sim\gamma_1$, the amplitude of the evanescent mode is nonzero at $a_2$ sites and vanishes at B$_2$ sites, and thus it cannot contribute to boundary condition of ZZ1, $F_{B2}(0,y)=0$.
As a result, the condition should be satisfied by the traveling mode alone, leading to the vanishing amplitude of the transmitted wave. 
For ZZ2, on the other hand, the condition $F_{a2}(0,y)=0$ is easily satisfied even for nonzero amplitude of the traveling mode because of the evanescent mode, leading to appreciable transmission.
\par
%
When $\gamma k/\gamma_1>\sqrt2$, the first excited conduction band crosses the Fermi energy and thus the second traveling mode opens for small incident angles between upward arrows.
In this case, the transmission probability is symmetric about $\theta=0$, causing no valley polarization, as has been discussed in the previous section.
For large incident angles (outside the downward arrows), there is no traveling mode in the bilayer graphene and therefore the transmission probability vanishes.
\par
%
Figure \ref{Fig:Charge_Density:_Armchair} shows transmission and reflection probabilities through the armchair boundary.
Inter-valley transmission and reflection probabilities are much smaller than the intra-valley probabilities when the electron density is sufficiently small, but slowly increase with energy and become comparable to intra-valley probabilities when the Fermi level reaches the bottom of the first excited conduction band.
\par
\begin{figure}[!t]
\begin{center}
\begin{center}
\includegraphics[width=7.0cm]{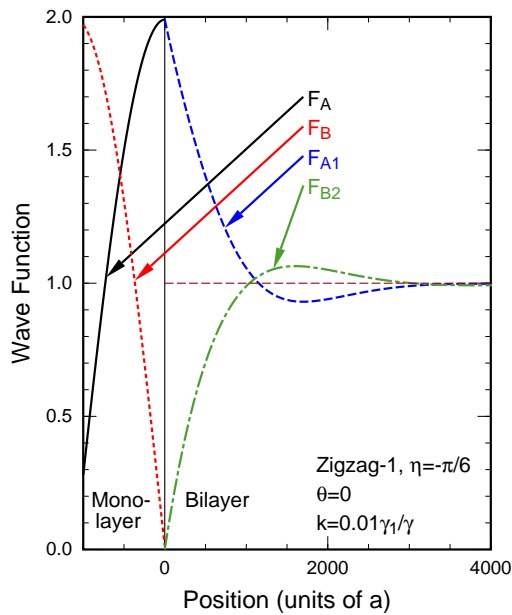}
\caption{(color online)
Calculated envelope function around a zigzag boundary ZZ1 ($\eta=-\pi/6$) for wave incident normal to the boundary.
The amplitude varies as a function of $x$ in the monolayer ($x<0$) due to interference of incident and reflected waves.
}
\label{Fig:Wave_Function}
\end{center}
\end{center}
\vspace{-0.50cm}
\end{figure}

%
Figure\ \ref{Fig:Wave_Function} shows some examples of the wave function as a function of position for a zigzag boundary ZZ1 with $\eta=-\pi/6$.
The energy is chosen to be sufficiently small, the incident angle $\theta=0$.
We note that $\tilde{F}_B(0)$ in the monolayer graphene becomes vanishingly small and consequently $\tilde{F}_A(0)\approx2$ in agreement with the discussion in Sec.\ IV.
Further, the boundary conditions (\ref{Eq:ZZ1_Envelop}) are satisfied by the presence of considerable amplitude of the evanescent mode.
In fact, the spatially-varying amplitude in the region $x>0$ mostly consists of the evanescent mode.
\par
%
\section{Discussion and Conclusion} \label{Sec:Discussion_and_Conclusion}
%
Explicit numerical calculations have been performed within the model of uniform charge density on both monolayer and bilayer regions.
In this model, the energy measured from the Dirac point can be slightly different between the layers when the electron density becomes nonzero (see Fig.\ \ref{Fig:Band_Alignment_and_Fermi_Lines}).
In actual systems, this may be realized by the presence of small potential variation in the vicinity of the boundary, which should be determined in a self-consistent manner.
The essential features of the results that envelope functions are well connected at the boundary and that strong valley polarization occurs due to the boundary transmission are expected to be independent of the presence of such small perturbations.
\par
%
%
We can also consider the case that the kinetic energy of the incident and transmitted waves is the same between two regions.
This is realized, for example, when a hot electron above the Fermi sea is injected.
The transmission is understood in the same manner, but there appears some significant difference because of the difference in the wave vector of the monolayer and bilayer, in particular, when the Fermi level lies in the vicinity of the Dirac point.
For a given wave vector $k$ in the monolayer, for example, the wave vector becomes $k'=\sqrt{\gamma_1k/\gamma}\gg k$ in the bilayer.
Because the wave-vector component $k_y$ parallel to the boundary is conserved, this leads to the focusing of the transmitted wave into the vertical direction, i.e., $|\theta'|< \arcsin |\varepsilon / \gamma_1|^{1/4}$, where $\theta'$ is the angle of the transmitted wave.
Further, we have $\kappa_x\approx k'\gg k_y$, showing that $k_y$ can be neglected in Eqs.\ (\ref{Eq:ApproximateZZ1}) and (\ref{Eq:ApproximateZZ2}).
Then, the transmission is nearly independent of the incident angle and the valley polarization is considerably reduced.
\par
\begin{figure}[!t]
\begin{center}
\includegraphics[width=8.50cm]{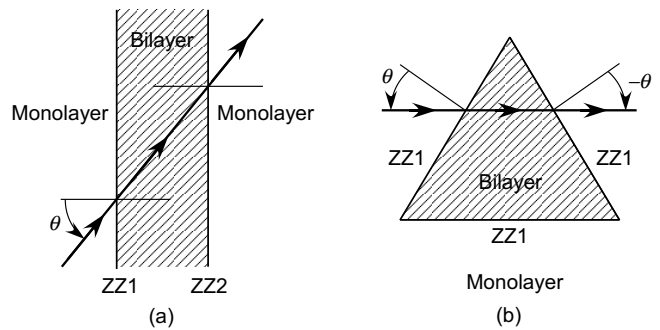}
\caption{Schematic illustration of ribbon and prism shaped region of bilayer graphene with zigzag boundary.
In a ribbon-shaped bilayer with zigzag structure, boundaries always consist of a pair of ZZ1 and ZZ2, but in the case of equilateral triangles, all boundaries consist of either ZZ1 or ZZ2.}
\label{Fig:Ribbon_and_Prism}
\end{center}
\vspace{-0.50cm}
\end{figure}

%
The valley polarization of waves transmitted through a single boundary is reduced when waves go through a ribbon-shaped narrow bilayer region sandwiched by monolayer graphenes, as shown in Fig.\ \ref{Fig:Ribbon_and_Prism}(a).
The reason lies in the cancellation at two parallel boundaries.
The time-reversal symmetry gives the relation that the transmission probability incident from the monolayer at the K point with angle $\theta$ is the same as that incident from the bilayer at the K' point in the reverse direction, i.e., $|t_{KK}^{\rm BM}(\theta)|^2=|t_{K'K'}^{\rm MB}(\theta)|^2$, where `BM' and `MB' stand for waves transmitted from monolayer to bilayer and from bilayer to monolayer, respectively.
Let us consider a hypothetical ribbon consisting only of ZZ1 boundary.
With the use of the symmetry $|t_{KK}^{\rm MB}(\theta)|^2=|t_{K'K'}^{\rm MB}(-\theta)|^2$, the total transmission probability through the bilayer ribbon is proportional to $|t_{KK}^{\rm MB}(\theta)|^2\times|t_{KK}^{\rm BM}(\theta)|^2=|t_{KK}^{\rm BM}(-\theta)|^2\times|t_{KK}^{\rm BM}(\theta)|^2$, when interference effects are neglected.
The result is independent of K and K' points.
\par
%
Actually, zigzag bilayer ribbons always consist of a pair of ZZ1 and ZZ2 boundaries as shown in Fig.\ \ref{Fig:Ribbon_and_Prism}(a), giving different amount of valley polarization.
Therefore, the cancellation is not complete and certain amount of valley polarization remains after transmission through a ribbon except in the vicinity of the Dirac point $|\varepsilon|\ll\gamma_1$, where the transmission probabilities across ZZ1 and ZZ2 are different only by factor two, leading to the complete cancellation.
This cancellation is reduced for two boundaries not parallel to each other and the polarization can be enhanced, for example, when waves go through a triangular-shape bilayer island formed in a monolayer graphene as shown in Fig.\ \ref{Fig:Ribbon_and_Prism}(b).
\par
%
%
Boundary conditions for edges of monolayer graphene with more general forms were discussed previously and boundaries were shown to be classified into either armchair or zigzag types.\cite{Akhmerov_and_Beenakker_2008a}
Similar considerations are likely to be possible in the present system.
For interfaces other than zigzag and armchair, however, the full boundary conditions require the presence of evanescent modes which are not described by states in the vicinity of the K and K' points given by Eqs.\ (\ref{Eq:Wave_Function_and_Envelopes_in_monolayer_graphene}) and (\ref{Eq:Wave_Function_and_Envelopes}).\cite{Akhmerov_and_Beenakker_2008a,Ando_and_Mori_1982a,Ando_et_al_1989a,Ando_and_Akera_1999a}
This problem is left for a future study.
\par
%
In conclusion, boundary conditions between monolayer and bilayer graphene have been obtained within an effective-mass scheme based on a tight-binding model.
Evanescent mode decaying exponentially away from the boundary plays an important role and as a result the traveling modes are strongly connected to each other between the monolayer and bilayer graphenes.
The transmission probability can be quite different between K and K' states for waves incident in oblique directions, resulting in significant valley polarization of waves transmitted through the boundary.
\par
%
\acknowledgments
%
This work was supported in part by Grant-in-Aid for Scientific Research on Priority Area ``Carbon Nanotube Nanoelectronics,'' by Grant-in-Aid for Scientific Research, and by Global Center of Excellence Program at Tokyo Tech ``Nanoscience and Quantum Physics'' from Ministry of Education, Culture, Sports, Science and Technology Japan.
\par
%
\appendix
%
\section{Low Energy Approximation} \label{Sec:Low_Energy_Approximation}
%
In order to understand boundary properties, the boundary condition (\ref{Eq:ZZ1_Envelop}) is examined in the low energy approximation $|\varepsilon|\ll \gamma_1$.
The envelope function in bilayer graphene consists of traveling wave $\tilde{\bf F}^K$ and evanescent wave ${\bf G}^K$.
Then, Eq.\ (\ref{Eq:ZZ1_Envelop}) becomes
%
\begin{equation}
\begin{array}{rl}
F^K_A&=\tilde{F}^K_{A1}+\alpha G^K_{A1}, \\
F^K_B&=\tilde{F}^K_{b1}+\alpha G^K_{b1}, \\
0&=\tilde{F}^K_{B2}+\alpha G^K_{B2},
\end{array}
\label{Eq:Boundary_Conditions_with_Evanescent_Mode}
\end{equation}
%
with coefficient $\alpha$.
In the low-energy regime, $\tilde{F}^K_{b1}$ can be replaced by $\tilde{F}^K_{B2}$ with the use of Eq.\ (\ref{Eq:Decomposition_Approx.}) and the evanescent wave (\ref{Eq:Evanescent_Mode}) is approximated by
%
\begin{equation}
 \pmatrix{G^{K}_{A1} \cr G^{K}_{b1} \cr G^{K}_{a2} \cr G^{K}_{B2} \cr }
= \pmatrix{ \displaystyle s {\kappa_x - k_y \over \kappa_x + k_y } \cr \displaystyle - i { \gamma (\kappa_x - k_y) \over \gamma_1 } \cr \displaystyle - i s {\gamma (\kappa_x - k_y ) \over \gamma_1 } \cr \noalign{\vspace{0.1250cm}} \displaystyle 1 \cr } .
\label{Evanescent_wave_approximation}
\end{equation}
%
Eliminating $\alpha$ in Eq.\ (\ref{Eq:Boundary_Conditions_with_Evanescent_Mode}), we immediately have Eq.\ (\ref{Conditions_for_Major_Components}).
\par
%
The envelope function in the monolayer side consists of incident wave in the direction $\theta$ and reflected wave in the direction $\pi-\theta$, i.e.,
%
\begin{equation}
\pmatrix{F^K_A(x) \cr \noalign{\vspace{0.0750cm}} F^K_B(x) \cr}
= \pmatrix{ \displaystyle  e^{-i\theta} \cr 1 \cr } e^{i k_x x} + r_{KK} \pmatrix{ \displaystyle - e^{i\theta} \cr 1 \cr } e^{- i k_x x} ,
\label{Eq:Incident_Wave}
\end{equation}
%
with reflection coefficient $r_{KK}$, where we use $k_x=sk\cos\theta$ and $k_y=sk\sin\theta$ in Eq. (\ref{Eq:Envelope_Function:_Monolayer}).
Under the condition of equal electron density in the monolayer and bilayer regions, the transmitted wave is written as
%
\begin{equation}
\pmatrix{\tilde{F}^K_{A1}(x) \cr \noalign{\vspace{0.0750cm}} \tilde{F}^K_{B2}(x)\cr}
 = f \pmatrix{ \displaystyle -s e^{-2i\theta} \cr 1 \cr } e^{i k_x x } ,
\label{Eq:Transmitted_Wave}
\end{equation}
%
with amplitude $f$.
\par
%
Upon substitution of Eqs.\ (\ref{Eq:Incident_Wave}) and (\ref{Eq:Transmitted_Wave}) in Eq.\ (\ref{Conditions_for_Major_Components}), and 
%
\begin{equation}
\kappa_x\approx\sqrt{|\varepsilon|\gamma_1/\gamma^2+k_y^2}\approx k\sqrt{1+\sin^2\theta}.
\label{Eq:ImaginaryWaveVector_Approx.}
\end{equation}
%
we have
%
\begin{eqnarray}
& \displaystyle f = 2\cos\theta \Big( -s e^{-2i\theta} -s {\kappa_x-k_y \over \kappa_x+k_y}\Big)^{-1} , & \label{Eq:1st_condition} \\
& \displaystyle r_{KK} = -1 - {\gamma k \over \gamma_1} \Big( s \cos \theta - i \sqrt{1 + \sin^2\theta} \Big) f . &
\end{eqnarray}
%
The transmission probability is given by $|t_{KK}|^2=2(\gamma k/\gamma_1)|f|^2$.
This satisfy the unitarity condition $|t_{KK}|^2\allowbreak+|r_{KK}|^2=1$ up to the lowest order in $\gamma k/\gamma_1$.
\par
%
\section{Armchair Boundary} \label{Sec:Armchair_Boundary}
%
An armchair boundary AC1, for example, shall be discussed in the vicinity of the Dirac point $|\varepsilon|\ll\gamma_1$.
After elimination of evanescent modes of the K and K' points, boundary condition for traveling modes becomes
%
\begin{equation}
\pmatrix{{\bf F}_A(0)\cr {\bf F}_B(0)\cr}\approx
\left.\pmatrix{\Gamma_{A A} & \Gamma_{A B}\cr
\Gamma_{BA} & \Gamma_{BB}\cr}
\pmatrix{\tilde{\bf F}_{A1}(x)\cr \tilde{\bf F}_{B2}(x)\cr}\right|_{x=0},
\label{Eq:032}
\end{equation}
%
with
%
\begin{eqnarray}
\Gamma_{AA} \! & = & \! \pmatrix{ 1 + \displaystyle {\kappa_x \!-\! k_y\over\kappa_x \!+\! k_y}{i\hat k_x \!-\! k_y\over 2\kappa_x} & \displaystyle - {\kappa_x \!-\! k_y\over\kappa_x \!+\! k_y}{i\hat k_x \!+\! k_y\over 2\kappa_x} \cr
\displaystyle - {\kappa_x \!+\! k_y\over\kappa_x \!-\! k_y}{i\hat k_x \!-\! k_y\over 2\kappa_x} & \displaystyle 1 + {\kappa_x \!+\! k_y \over \kappa_x \!-\! k_y}{i\hat k_x \!+\! k_y\over 2\kappa_x} \cr } , \nonumber \\
\Gamma_{AB} \! & = & \! \displaystyle - {s\over 2} \pmatrix{
\displaystyle {\kappa_x \!-\! k_y \over \kappa_x} & \displaystyle {\kappa_x \!-\! k_y \over \kappa_x} \cr
\displaystyle {\kappa_x \!+\! k_y \over \kappa_x} & \displaystyle {\kappa_x \!+\! k_y \over \kappa_x} \cr } , \nonumber \\
\Gamma_{BA} \! & = & \! \displaystyle {s\gamma\over 2\gamma_1} \pmatrix{
\displaystyle {\kappa_x \!-\! k_y \over \kappa_x} (\hat k_x \!+\! i k_y) & \displaystyle \!\! -{\kappa_x \!-\! k_y \over \kappa_x} (\hat k_x \!-\! i k_y) \cr
\displaystyle \!\! - {\kappa_x \!+\! k_y \over \kappa_x} (\hat k_x \!+\! i k_y) & \displaystyle {\kappa_x \!+\! k_y \over \kappa_x} (\hat k_x \!-\! i k_y) \cr } , \nonumber \\
\Gamma_{BB} \! & = & \! \displaystyle -{\gamma\over \gamma_1}
\pmatrix{ \hat k_x \!-\! i k_y\!\!\!\!\!\! &0 \cr
0 &\!\!\!\!\!\!\hat k_x \!+\! i k_y \cr }
+ {i\over 2} {\gamma\over \gamma_1} {\kappa_x^2 \!-\! k_y^2\over\kappa_x}
\pmatrix{1 & 1 \cr 1 & 1 \cr} , \nonumber \\
\end{eqnarray}
%
and
%
\begin{equation}
{\bf F}_A(x)=\pmatrix{{F}^K_A(x)\cr {F}^{K'}_A(x)\cr}, \ \ \mbox{etc}.
\label{Eq:Vector of Envelope}
\end{equation}
%
In the limit $|\varepsilon|/\gamma_1\rightarrow0$, they are reduced to
%
\begin{equation}
\begin{array}{ll}
{\bf F}_A(0)&\! \displaystyle \approx \Gamma_{AA} \tilde{\bf F}_{A1}(x) \big|_{x=0} +\Gamma_{AB} \tilde{\bf F}_{B2}(x) \big|_{x=0} , \\
{\bf F}_B(0) &\! \approx 0 .
\end{array}
\label{Eq:Armchair epsilon=0}
\end{equation}
%
\par
%
This shows that $r_{KK}=-1$ and $r_{K'K}=0$ for electron wave incident from the K valley at $k=0$, the same as for zigzag boundaries.
With the increase of $k$, the transmission increases in proportion to $k$ and its amplitude can be estimated using the first equation of (\ref{Eq:Armchair epsilon=0}).
Because of the presence of off-diagonal elements in $\Gamma_{AA}$ and $\Gamma_{AB}$, inter-valley mixing occurs at the armchair boundary in proportion to $k$.
After some manipulations, the amplitude $f$ transmitted into K valley and $f'$ into the K' valley become
%
\begin{eqnarray}
f\! & = & \! -{s\over 4}e^{2 i \theta}\cos\theta\left(5-e^{2 i \theta}-2 i s e^{i \theta}\sqrt{1+\sin^2\theta}\right) , \nonumber \\
f' \! & = & \! \displaystyle {1\over 2}f_{0}e^{i (-2\theta+\alpha)} , \\
\alpha \! & = & \! \displaystyle s \tan^{-1} {\cos\theta\sqrt{1+\sin^2\theta}\over\sin^2\theta} . \nonumber
\label{Eq:Ratio_of_wave:_AC}
\end{eqnarray}
%
For the K point, the transmission probability is proportional to $k|f|^2$ which takes maximum at $\theta\approx 0.174\pi$, and for the K' point $k|f'|^2$ which takes maximum at $\theta_0$.
Analysis of the above equations reveals that inter-valley mixing is 1/5 of the transmission probability within valley for perpendicularly incident wave ($\theta=0$).
The total probability is given by the sum of them and maximum transmission occurs at $\theta\approx 0.179\pi$.
\par
%
\section{Edge States} \label{Sec:Edge_States}
%
As in edges of monolayer graphene,\cite{Fujita_et_al_1996a,Nakada_et_al_1996a} there can be edge states localized along a boundary between the monolayer and bilayer graphene.
An edge state consists of evanescent modes exponentially decaying in the negative $x$ direction in the monolayer and those decaying in the positive $x$ direction in the bilayer.
In the following, we shall confine ourselves to the case of vanishing electron density in both monolayer and bilayer regions.
\par
%
In monolayer graphene occupying half space $x<0$, a relevant evanescent mode with energy $\varepsilon$ and $k_y$ in the range $|\gamma k_y| > |\varepsilon|$ has imaginary wave vector $i\kappa$, with
%
\begin{equation}
\gamma \kappa = - \sqrt{(\gamma k_y)^2 - \varepsilon^2 } ,
\end{equation}
%
and the wave function ${\bf G}_-^K e^{-\kappa x + i k_y y}$ for the K point, with
%
\begin{equation}
{\bf G}_-^K = \pmatrix{ G_A^{K} \cr G_B^{K} \cr } \equiv \pmatrix{ \displaystyle + s_\varepsilon s_y \sqrt{\Big| { \kappa - k_y \over 2 k_y } \Big|} \cr \noalign{\vspace{0.10cm}} \displaystyle i \sqrt{\Big| { \kappa + k_y \over 2 k_y } \Big|} \cr } , 
\end{equation}
%
where $s_\varepsilon$ and $s_y$ denote the sign of $\varepsilon$ and $k_y$, respectively.
The wave function for the K' point is obtained by replacing $k_y$ with $-k_y$.
\par
%
In bilayer graphene lying in the region $x>0$, we can have two evanescent modes with wave vector
%
\begin{eqnarray}
& \gamma \kappa_j = + \sqrt{(\gamma k_y)^2 - \varepsilon^2 + s_j |\varepsilon| \gamma_1 } , & \\
& s_j = \left\{ \begin{array}{ll} -1 & (j=1); \\ +1 & (j=2), \end{array} \right. &
\end{eqnarray}
%
and wave function ${\bf G}_{+j}^K e^{-\kappa_j x + i k_y y}$, with
%
\begin{equation}
{\bf G}_{+j}^K
= \pmatrix{ G_{A1}^{Kj} \cr \noalign{\vspace{0.10cm}} G_{B1}^{Kj} \cr \noalign{\vspace{0.10cm}} G_{A2}^{Kj} \cr \noalign{\vspace{0.10cm}} G_{B2}^{Kj} \cr }
\equiv {1\over 2} \pmatrix{ \displaystyle - { s_j s_\varepsilon \gamma ( \kappa_{j} \!-\! k_y ) \over \sqrt{ (\gamma k_y)^2 \!+\! {1\over 2} s_j |\varepsilon| \gamma_1 } } \cr \noalign{\vspace{0.10cm}} \displaystyle i { s_j |\varepsilon| \over \sqrt{ (\gamma k_y)^2 \!+\! {1\over 2} s_j |\varepsilon| \gamma_1 } } \cr \noalign{\vspace{0.10cm}} \displaystyle - i {\varepsilon \over \sqrt{ (\gamma k_y)^2 \!+\! {1\over 2} s_j |\varepsilon| \gamma_1 } } \cr \noalign{\vspace{0.10cm}} \displaystyle {\gamma ( \kappa_{j} \!+\! k_y ) \over \sqrt{ (\gamma k_y)^2 \!+\! {1\over 2} s_j |\varepsilon| \gamma_1 } } \cr } .
\label{Eq:Two_Evanescent_Modes}
\end{equation}
%
These evanescent modes exist in the region $|\gamma k_y| > \sqrt{\varepsilon^2 - s_j |\varepsilon|\gamma_1}$.
Therefore, there are no traveling modes in both monolayer and bilayer graphenes in the region
%
\begin{eqnarray}
|\gamma k_y| > \sqrt{\varepsilon^2 + |\varepsilon|\gamma_1} .
\label{Eq:Evanescent_Region_in_Bilayer}
\end{eqnarray}
%
Note that ${\bf G}_{+2}^K$ is the same as Eq.\ (\ref{Eq:Evanescent_Mode}).
\par
%
Edge states localized near the boundary ($x=0$) have the wave function
%
\begin{equation}
{\bf G}({\bf r}) = \left\{\begin{array}{ll} \displaystyle \sum_{v=K,K'} \alpha_v {\bf G}_-^v e^{-\kappa x + i k_y y} & (x<0); \\ \noalign{\vspace{0.10cm}} \displaystyle \sum_{v=K,K'} \sum_{j=1,2} \beta_{vj} {\bf G}_{+j}^v e^{-\kappa_{j} x + i k_y y} & (x>0), \end{array} \right.
\end{equation}
%
with appropriate coefficients $\alpha_v$ and $\beta_{vj}$.
More explicitly, for boundary ZZ1, we have
%
\begin{equation}
\pmatrix{ G_{A1}^{v1} & G_{A1}^{v2} & G_A^{v} \cr
i G_{B1}^{v1} & i G_{B1}^{v2} & i G_B^{v} \cr
G_{B2}^{v1} & G_{B2}^{v2} & 0 \cr }
\pmatrix{ \beta_{v1} \cr \beta_{v2} \cr - \alpha_{v} \cr }
= 0 ,
\end{equation}
%
where we have multiplied imaginary unit $i$ in such a way that the coefficient matrix becomes real.
For boundary ZZ2, we have
%
\begin{equation}
\pmatrix{ G_{A1}^{v1} & G_{A1}^{v2} & G_A^{v} \cr
i G_{B1}^{v1} & i G_{B1}^{v2} & i G_B^{v} \cr
i G_{A2}^{v1} & i G_{A2}^{v2} & 0 \cr }
\pmatrix{ \beta_{v1} \cr \beta_{v2} \cr - \alpha_v \cr }
= 0 ,
\end{equation}
%
For AC1, we have
%
\begin{eqnarray}
& \!\!\! \pmatrix{ G_{A1}^{K1} & G_{A1}^{K2} & G_A^{K} & \!\! 0 & 0 & 0 \cr
i G_{B1}^{K1} & \! i G_{B1}^{K2} \! & \! i G_B^{K} \! & \!\! 0 & 0 & 0 \cr
G_{B2}^{K1} & G_{B2}^{K2} & 0 & \!\! G_{B2}^{K'1} & G_{B2}^{K'2} & 0 \cr
0 & 0 & 0 & \!\! G_{A1}^{K'1} & G_{A1}^{K'2} & G_A^{K'} \! \cr
0 & 0 & 0 & \!\!\! i G_{B1}^{K'1} & \! i G_{B1}^{K'2} & \! i G_B^{K'} \! \cr
\! - i G_{A2}^{K1} & \!\!\! - i G_{A2}^{K2} \! & 0 & \!\!\! i G_{A2}^{K'1} & \!\ i G_{A2}^{K'2} & 0 \cr } \!\!
\pmatrix{ \beta_{K1} \cr \beta_{K2} \cr - \alpha_K \cr \beta_{K'1} \cr \beta_{K'2} \cr - \alpha_{K'} \cr } \!\!\!\! & \nonumber \\
& \qquad\qquad\qquad\qquad\qquad\qquad\qquad\qquad = 0 . &
\end{eqnarray}
%
The determinant of the coefficient matrix remains nonzero in the energy range satisfying Eq.\ (\ref{Eq:Evanescent_Region_in_Bilayer}) and vanishes at $\varepsilon=0$.
Therefore, edge states can be present only at $\varepsilon=0$.
\par
%
Let us consider the special case $\varepsilon=+0$ or $\varepsilon=-0$.
In the monolayer region, we have $\kappa=-k_y$ for $k_y>0$, $\kappa=k_y$ for $k_y<0$, and therefore the evanescent mode becomes
%
\begin{equation}
{\bf G}_-^K = \left\{ \begin{array}{ll} \pmatrix{ 1 \cr 0 \cr } & (k_y>0); \\ \noalign{\vspace{0.10cm}} \pmatrix{ 0 \cr i \cr } & (k_y<0), \end{array} \right.
\end{equation}
%
where we have multiplied an appropriate phase factor.
The wave function for the K' point is obtained by replacing $k_y$ with $-k_y$.
\par
%
In the bilayer region, on the other hand, we have $\kappa_j=|k_y|$ for both $j=1$ and 2, and consequently ${\bf G}_{+j}^K$ becomes the same between $j=1$ and $2$.
In order to obtain two independent evanescent modes we expand ${\bf G}_{+j}^K(x) \equiv {\bf G}_{+j}^Ke^{-\kappa_jx}$ in terms of $\delta=|\varepsilon|\gamma_1/(\gamma k_y)^2$,
%
\begin{equation}
{\bf G}_{j+}^K(x) = {\bf G}_{j+}^K(x)^{(0)} + \delta \, {\bf G}_{j+}^K(x)^{(1)} + O(\delta^2) ,
\end{equation}
%
with
%
\begin{equation}
{\bf G}_{+j}^K(x)^{(0)} = \pmatrix{ \displaystyle - s_j s_\varepsilon {1 \!-\! s_y \over 2} \cr 0 \cr 0 \cr \displaystyle {1\!+\! s_y \over 2} \cr } e^{-|k_y|x } ,
\end{equation}
%
and
%
\begin{eqnarray}
& \displaystyle {\bf G}_{+j}^K(x)^{(1)} = - {1\over 4} s_j \pmatrix{ \displaystyle - s_j s_\varepsilon {1 \!-\! s_y \over 2} \cr \noalign{\smallskip} 0 \cr\noalign{\smallskip}  0 \cr \noalign{\smallskip} \displaystyle {1\!+\! s_y \over 2} \cr } ( 1 \!+\! 2|k_y| x ) e^{-|k_y|x } & \nonumber \\
& \displaystyle - {1\over 2} s_j \pmatrix{\displaystyle {s_j s_\varepsilon \over 2} \cr \displaystyle - \i {\gamma|k_y|\over \gamma_1 } \cr \displaystyle \i s_j s_\varepsilon {\gamma|k_y|\over \gamma_1} \cr \displaystyle - {1\over 2} \cr }  e^{-|k_y|x } . &
\end{eqnarray}
%
Then, two independent modes can be written as ${\bf G}_1^K(x)$ and ${\bf G}_2^K(x)$ with ${\bf G}_1^K(x)={\bf G}_{+1}^K(x)^{(0)}$ and
%
\begin{equation}
{\bf G}_2^K(x) = \left\{ \begin{array}{ll} \displaystyle \sum_{j=1,2} s_j {\bf G}_{+j}^K(x)^{(1)} & (k_y>0); \\ \noalign{\vspace{0.10cm}} \displaystyle \sum_{j=1,2} {\bf G}_{+j}^K(x)^{(1)} & (k_y<0). \end{array} \right.
\end{equation}
%
Therefore, we have for $k_y>0$
%
\begin{eqnarray}
& {\bf G}_1^K(x) = \pmatrix{ 0 \cr 0 \cr 0 \cr 1 \cr } e^{-|k_y|x} , & \\
& {\bf G}_2^{K}(x) = \pmatrix{ 0 \cr \noalign{\smallskip} \displaystyle i {\gamma|k_y|\over \gamma_1} \cr \noalign{\smallskip} 0 \cr \noalign{\smallskip} \displaystyle {1\over 2} \!-\! |k_y| x \cr } e^{-|k_y|x} , &
\end{eqnarray}
%
and for $k_y<0$
%
\begin{eqnarray}
& {\bf G}_1^K(x) = \pmatrix{ 1 \cr 0 \cr 0 \cr 0 \cr } e^{-|k_y|x} , & \\
& {\bf G}_2^{K}(x) = \pmatrix{ \displaystyle {1\over 2} \!-\! |k_y| x \cr \noalign{\vspace{0.10cm}} 0 \cr \displaystyle i {\gamma|k_y|\over \gamma_1} \cr \noalign{\smallskip} 0 \cr } e^{-|k_y|x} . &
\end{eqnarray}
%
The wave functions for the K' point are again obtained by replacing $k_y$ with $-k_y$, i.e., ${\bf G}_j^{K'}(x;k_y)={\bf G}_j^{K}(x;-k_y)$.
\par
%
Therefore, we have for $k_y>0$
%
\begin{eqnarray}
{\rm ZZ1:} && \!\!\!\! \bigg\{ \begin{array}{rl} K: & \alpha_K = \beta_{K1} = \beta_{K2} = 0 , \\ K': & \alpha_{K'} = 0 , \quad \beta_{K'1} + {1\over2} \beta_{K'2} = 0 , \end{array} \\
{\rm ZZ2:} && \!\!\!\! \bigg\{ \begin{array}{rl} K: & \alpha_K = \beta_{K2} = 0 , \\ K': & \alpha_{K'} = \beta_{K'1} = \beta_{K'2} = 0 , \end{array} \\
{\rm AC1:} && \!\!\!\! \bigg\{ \begin{array}{rl} \alpha_K = \beta_{K1} = \beta_{K2} = 0 , \\ \alpha_{K'} = \beta_{K'1} = \beta_{K'2} = 0 , \end{array} 
\end{eqnarray}
%
and for $k_y<0$
%
\begin{eqnarray}
{\rm ZZ1:} && \!\!\!\! \bigg\{ \begin{array}{rl} K: & \alpha_K = 0 , \quad \beta_{K1} + {1\over 2} \beta_{K2} = 0 , \\ K': & \alpha_{K'} = \beta_{K'1} = \beta_{K'2} = 0 , \end{array} \\
{\rm ZZ2:} && \!\!\!\! \bigg\{ \begin{array}{rl} K: & \alpha_K = \beta_{K1} = \beta_{K2} = 0 , \\ K': & \alpha_{K'} = \beta_{K'2} = 0 , \end{array} \\
{\rm AC1:} && \!\!\!\! \bigg\{ \begin{array}{rl} \alpha_K = \beta_{K1} = \beta_{K2} = 0 , \\ \alpha_{K'} = \beta_{K'1} = \beta_{K'2} = 0 . \end{array}
\end{eqnarray}
%
There is no edge state in the armchair boundary.
\par
%
In the case of boundary ZZ1, we have a single edge state at the K' point for $k_y>0$ and one at the K point for $k_y<0$.
The wave function of these states is completely localized in the bilayer region and is given by 
%
\begin{equation}
{\bf G}^{K}(x) = \pmatrix{ |k_y| x \cr \noalign{\smallskip} 0 \cr \noalign{\smallskip} \displaystyle - i {\gamma |k_y| \over \gamma_1 } \cr \noalign{\smallskip} 0 \cr } e^{-|k_y|x}  \quad (k_y<0) .
\end{equation}
%
The wave function for the K' point ($k_y>0$) is also given by the above equation.
\par
%
In the case of boundary ZZ2, on the other hand, we have a single edge state at the K point for $k_y>0$ and one at the K' point for $k_y<0$.
The wave function is completely localized in the bilayer region and is given by 
%
\begin{equation}
{\bf G}^{K}(x) = \pmatrix{ 0 \cr 0 \cr 0 \cr 1 \cr } e^{-|k_y|x} \quad (k_y>0).
\end{equation}
%
The wave function for the K' point ($k_y<0$) is again given by the same expression.
These results are summarized in Table \ref{Tab:Number_of_Edge_States}.
\par
%
For $j\!=\!2$ and at $|\varepsilon|\!=\!\gamma_1$, we have $\kappa_{2}\!=\!|k_y|$, giving $G_{B2}^{K2}\!=\!0$ for $k_y\!<\!0$ and $G_{B2}^{K'2}\!=\!0$ for $k_y\!>\!0$ in Eq.\ (\ref{Eq:Two_Evanescent_Modes}).
Other elements of ${\bf G}_{j}^K$ and ${\bf G}_{j}^{K'}$ all remain nonzero.
For the boundary ZZ1, therefore, the boundary condition $F_{B2}(0,y)\!=\!0$ is satisfied and traveling modes in the monolayer can be connected only to the evanescent mode.
It is easy to show that this evanescent mode cannot be connected to the evanescent mode in the monolayer and therefore cannot form a pure edge state.
However, we have perfect reflection at $|\varepsilon|\!=\!\gamma_1$ for $-\gamma_1\!<\!\gamma k_y\!<\!0$ at the K point and for $0\!<\!\gamma k_y\!<\!\gamma_1$ at the K' point.
\par
%
This perfect reflection is closely related to the vanishing transmission probability at $\varepsilon=\gamma_1$ for ZZ1 shown in Fig.\ \ref{Fig:Transmission_vs_Incident_Angle2}.
In fact, when the Fermi level lies at $\gamma_1$ under the condition that the electron density is the same between the monolayer and bilayer graphenes, i.e., $k=\sqrt2\gamma_1/\gamma$, the reflection coefficient for wave incident from the monolayer side is calculated as
%
\begin{equation}
r_{KK} = { \cos\theta - i ( 1 \!-\! 2\sqrt2) \sin\theta \over \cos\theta + i ( 1 \!-\! 2\sqrt2) \sin\theta } \quad (\theta>0) .
\end{equation}
%
\par
%

%
\vspace{0.150cm}
\hrule
\vspace{0.150cm}
\rightline{File: \jobname.tex (\today)}
%
\enddocument